\documentclass{emulateapj}

\usepackage{hyperref}

\hypersetup{
    bookmarks=true,         
    unicode=false,          
    pdftoolbar=true,        
    pdfmenubar=true,        
    pdffitwindow=true,     
    pdfstartview={FitH},    
    pdftitle={Search for MP Stars},    
    pdfauthor={vmplacco},     
    pdfsubject={Astronomy},   
    pdfcreator={dvipdf},   
    pdfproducer={dvipdf}, 
    pdfkeywords={metal poor stars, carbon enhancement, HES},
    pdfnewwindow=true,      
    colorlinks=true,       
    linkcolor=red,          
    citecolor=blue,        
    filecolor=magenta,      
    urlcolor=cyan,           
    breaklinks=true,
    linktocpage
}

\newcommand{\metal}{[Fe/{}H]}
\newcommand{\cfe}{[C/{}Fe]}
\newcommand{\abund}[2]{[#1/{}#2]}
\newcommand{\hes}{{HES}}

\newcommand{\cn}{{CN}}

\newcommand{\cemp}{{CEMP}}
\newcommand{\ctwo}{{C$_2$}}

\newcommand{\jk}{({J$-$K})$_0$}
\newcommand{\bvz}{({B$-$V})$_0$}

\newcommand{\gpe}{{GPE}}
\newcommand{\egp}{{EGP}}

\newcommand{\nc}{\newcommand}
\nc{\teff}{$T_{\rm eff}$\,}  
\nc{\logg}{log\,$g$\,}

\shorttitle{Search for MP Stars}
\shortauthors{Placco et al.}

\begin{document}

\title{Searches for Metal-Poor Stars from the Hamburg/ESO Survey using the CH G-band}

\author{Vinicius M. Placco}
\affil{Departamento de Astronomia - Instituto de Astronomia, Geof\'isica e Ci\^encias Atmosf\'ericas, Universidade de S\~ao Paulo, S\~ao Paulo, SP 05508-900, Brazil}
\email{vmplacco@astro.iag.usp.br}

\author{Catherine R. Kennedy, Timothy C. Beers}
\affil{Department of Physics \& Astronomy and JINA: Joint Institute for Nuclear Astrophysics, Michigan State University, East Lansing, MI 48824, USA}

\author{Norbert Christlieb}
\affil{Zentrum f\"ur Astronomie der Universit\"at Heidelberg, Landessternwarte, K\"onigstuhl 12, 69117, Heidelberg, Germany}

\author{Silvia Rossi}
\affil{Departamento de Astronomia - Instituto de Astronomia, Geof\'isica e Ci\^encias Atmosf\'ericas, Universidade de S\~ao Paulo, S\~ao Paulo, SP 05508-900, Brazil}

\author{Thirupathi Sivarani}
\affil{Indian Institute of Astrophysics, 2nd block, Koramangala, Bangalore 560034, India}

\author{Young Sun Lee}
\affil{Department of Physics \& Astronomy and JINA: Joint Institute for Nuclear Astrophysics, Michigan State University, East Lansing, MI 48824, USA}

\author{Dieter Reimers}
\affil{Hamburger Sternwarte, Universit\"at Hamburg, Gojenbergsweg 112, 21029 Hamburg, Germany}

\author{Lutz Wisotzki}
\affil{Astrophysical Institute Potsdam, An der Sternwarte 16, 14482 Potsdam, Germany}

\accepted{for publication in AJ – September 18, 2011}

\begin{abstract}

We describe a new method to search for metal-poor candidates from the
Hamburg/ESO objective-prism survey (\hes) based on identifying stars with
apparently strong CH G-band strengths for their colors. The
hypothesis we exploit is that large over-abundances of carbon are common among
metal-poor stars, as has been found by numerous studies over the past two
decades. The selection was made by considering two line indices in the 4300\,
{\AA} region, applied directly to the low-resolution prism spectra. This work
also extends a previously published method by adding bright sources to the
sample. The spectra of these stars suffer from saturation effects, compromising
the index calculations and leading to an undersampling of the brighter
candidates. A simple numerical procedure, based on available photometry, was
developed to correct the line indices and overcome this limitation. Visual
inspection and classification of the spectra from the \hes{} plates yielded a
list of 5,288 new metal-poor (and by selection, carbon-rich) candidates, which
are presently being used as targets for medium-resolution spectroscopic
follow-up. Estimates of the stellar atmospheric parameters, as well as carbon
abundances, are now available for 117 of the first candidates, based on
follow-up medium-resolution spectra obtained with the SOAR 4.1m and Gemini 8m
telescopes. We demonstrate that our new method improves the metal-poor star
fractions found by our pilot study by up to a factor of three in the same
magnitude range, as compared with our pilot study based on only one CH G-band
index. Our selection scheme obtained roughly a 40\% success rate for
identification of stars with \metal~$< -1.0$; the primary contaminant is
late-type stars with near solar abundances and, often, emission line cores that
filled in the Ca\,{\sc{ii}} K line on the prism spectrum. Because the selection
is based on carbon, we greatly increase the numbers of known \cemp{} stars from
the HES with intermediate metallicities $-2.0 <$~\metal~$< -1.0$, which previous
survey efforts undersampled. There are eight newly discovered stars with
\metal~$< -3.0$ in our sample, including two with
\metal~$< -3.5$.

\end{abstract}


\section{Introduction}
\label{intro}

The interest in stars with metallicities lower than about ten times the solar
value (\metal\footnote{\abund{A}{B} = $log(N_A/{}N_B)_{\star} - log(N_A/{}N_B)
_{\odot}$, where $N$ is the number density of atoms of a given element, and the
indices refers to the star ($\star$) and the Sun ($\odot$).}$<$$-$1.0) has
greatly intensified in the past few decades, due to their utility as probes of
the chemical history and kinematics of the stellar populations of the Galaxy and
the nucleosynthesis pathways explored by the first generations of stars. Among
the most interesting stars for such applications are those with the lowest
possible metallicity. These chemically most primitive stars are expected to
provide one of the best windows on chemical evolution during the first 0.5-1 Gyr
following the Big Bang. Such stars are extraordinarily rare (only four stars
with \metal$< -4.5$ are known at present), and require inspection of very large
samples of candidates to identify them. The key is being able to distinguish
likely metal-poor stars from among the overwhelmingly greater numbers of solar
or near-solar abundance stars in the Galaxy.

Beginning roughly 40 years ago with the pioneering work of \citet{bond1970,
bond1980}, \citet{slettebak1971}, and \citet{bm1973}, photographic
objective-prism techniques were shown to be efficient sieves for the
identification of large numbers of metal-poor (and chemically peculiar) stars
for further study. These efforts were expanded by the HK Survey of Beers,
Preston, \& Shectman \citep{beers1985,beers1992}, and later by the Hamburg/{}ESO
Survey \citep[\hes{}; ][]{reimers1997, christlieb2003}, to include fainter stars
that explore farther into the halo system of the Galaxy where the largest
numbers of metal-poor stars have been found. Both of these surveys sought to
identify metal-poor candidates by visual (HK Survey) or digitized (\hes)
scans of the prism plates, looking for stars with weak or absent lines of Ca\,
{\sc{ii}} in low-resolution spectra. Star-by-star follow-up medium-resolution
spectroscopy for over 15,000 candidates from these two surveys consumed large
amounts of 1.5m to 4m telescope time over the past two decades, but yielded over
3000 very metal-poor (VMP; \metal\ $ < -2.0$) stars, and several hundred extremely
metal-poor (EMP; \metal\ $< -3.0$) stars. See \citet{beers2005} for a more
complete history of these efforts.

More recently, the Sloan Digital Sky Survey \citep[SDSS;][]{york2000}, in
particular the sub-surveys known as the Sloan Extension for Galactic
Understanding and Exploration \citep[SEGUE;][]{yanny2009} and SEGUE-2 
\citep{rockosi2011} were able to obtain medium-resolution spectroscopy for almost
500,000 stars, including numerous metal-poor candidates selected on the basis of
their broadband $ugriz$ colors. These efforts have yielded the identification of
over 25,000 VMP stars and on the order of 1000 EMP stars. Even larger samples
are expected to come from the Chinese LAMOST \citep[Large sky Area Multi-Object
fiber Spectroscopic Telescope;][]{zhao2006} and the Australian SkyMapper 
\citep{keller2007} projects.  

One of the limitations of the metal-poor stars identified by the SDSS is that
the sample has a bright limit of $g \sim 14.0$, set by the saturation level of
the photometric scans used to identify candidates in the first place. To the
extent that SDSS imaging is used for targeting LAMOST targets, this limit will
similarly apply to that survey. Stars brighter than 14th magnitude require far
less 6.5m-10m telescope time in order to obtain the high-resolution spectroscopy
that enables detailed understanding of their abundance patterns, and hence are
greatly desired. The SkyMapper project will not suffer this limitation, since it
includes very short exposure times in its planned cadence. However, as we show
below, it is possible to use the already available \hes{} prism spectra and
2MASS \citep[Two Micron All Sky Survey; ][]{2mass} near-IR photometry to improve
our ability to identify metal-poor candidates that include brighter stars.  

Our technique is based on the observational fact that, at very low metallicity,
an increasing fraction of very metal-poor stars exhibit strong over-abudances of
carbon. The great majority of stars more metal-rich than \metal\ $= -1.0$ exhibit
carbon-to-iron ratios, \cfe, that track closely with \metal. However, below
\metal\ $= -2.0$, on the order of 20\% of stars have \cfe $> + 1.0$ 
\citep{lucatello2006}. The fraction of so-called carbon-enhanced metal-poor 
\citep[\cemp,][]{beers2005} stars in the HK survey and the HES rises to 30$\%$ for
\metal\ $<-$3.0, 40$\%$ for \metal\ $<-$3.5, and 75$\%$ for \metal\ $<-$4.0
\citep[accounting for the recently discovered {\it non}-carbon-enhanced star with \metal $\sim
-5.0$, identified by][from among metal-poor stars in the SDSS]{caffau2011}.

\citet{carollo2011} have argued that the increase in the frequency of \cemp{}
stars with declining metallicity is due to the fact that the outer-halo
component of the Galaxy possesses about twice the fraction of \cemp{} stars
relative to carbon-normal stars {\emph{at a given low metallicity}} than the
inner-halo component. The observed correlation between metallicity and
CEMP fraction is a manifestation of the lower metallicity of
outer-halo stars, which begin to dominate halo samples at low
abundance\footnote{\citet{carollo2007,carollo2010} report that the peak of the
inner-halo metallicity distribution function occurs at \metal~= $-1.6$, while
that of the outer halo falls at \metal~= $-2.2$.}. This idea can also account
for the observed increase in the fraction of \cemp{} stars at a given
metallicity as a function of height above the Galactic plane
\citep{frebel2006,carollo2011}, and may have influenced previous reports of
less than a 20\% fraction of \cemp{} stars in the HES at very low metallicity
\citep{cohen2005,frebel2006,frebel2009}.


The \citet{aoki2007} study of some 26 CEMP stars reveals stars 
with \metal\ $< -$2.2 that do not exhibit evidence for the operation of the 
s-process (the CEMP-no stars). Furthermore, in two of the most iron-deficient 
stars known today, HE~0557-4840 \citep[\metal = $-$4.8;][]{norris2007}, and HE~0107-5240 
\citep[\metal = $-$5.3;][]{christlieb2002} no neutron-capture elements have been detected; in 
HE~1327-2326 \citep[\metal = $-$5.4;][]{frebel2005}, Sr has been detected, but the upper 
limit for Ba indicates that the Sr in this star was not produced in the 
main s-process. Besides that, three of the four known ultra metal-poor (UMP)
and hyper metal-poor (HMP) stars have huge over-abundances of CNO, up 
to several thousand times the solar ratios. Thus, it seems evident that 
the mechanisms by which carbon (and similarly N and O) has been enhanced 
in metal-deficient stars are likely to be much more diverse than can be 
accounted for by any single process. Corroborating this hypothesis, 
\citet{cooke2011} reports on the identification of a high redshift, 
extremely metal-poor (\metal$\sim-$3.0) Damped Lyman Alpha (DLA) system 
with an observed pattern of CNO that closely resembles that of the UMP stars.

These results immediately suggest that at some early time in the Universe a
significant amount of carbon was produced, by one or more of the following
sources: (1) a primordial mechanism from massive, zero-metallicity, rapidly
rotating stellar progenitors \citep{hirschi2006,meynet2006,meynet2010}, (2)
production by ``faint supernovae'', which eject large amounts of CNO during
their explosions \citep{umeda2005,kobayashi2011}, or (3) production of carbon by
stars of intermediate mass, which can be prodigious manufacturers of carbon
during their AGB stages, followed by mass transfer to a surviving lower-mass
companion. It remains possible that all three sources have played a role.


\citet{placco2010} (hereafter Paper I) shows that it is possible to
search for metal-poor stars based on the premise that a large fraction of them
will also be carbon enhanced, and that fraction will increase at the lowest
metallicities. They reported that, by making a selection based on a new line
index for the CH G-band (\gpe{} -- applied to the low-resolution HES prism
spectra) as a function of 2MASS $(J-K)_0$ color, more than 65\% of their
candidates had \metal~$< -1.0$, while some 23\% had \metal~$< -2.0$, based on
medium-resolution spectroscopic follow-up. Not surprisingly, many of the
candidates turned out to be \cemp{} stars as well. The technique clearly works,
and what remains to be done is to refine this approach and improve its selection
efficiency by eliminating sources of uninteresting candidates entering the
sample.

The present study aims to extend the work initiated in Paper I by introducing an
auxiliary line index for the CH G-band, one that we show below improves the
fractions of bona-fide \cemp{} stars selected for follow-up spectroscopy by
roughly a factor of two compared to our initial effort. In addition, we have
increased our magnitude limit for medium-resolution observations and included
the brighter stars from the \hes, developing a correction scheme to reduce
saturation effects on the line indices. By refining our methods to search for
carbon-enhanced stars, we expect to continue increasing the number of known \cemp{}
stars with intermediate metallicities, as well as explore regimes with
\metal\ $<-3.0$, where carbon plays a major role in describing the chemical
evolution processes and formation scenarios of our Galaxy.

This paper is outlined as follows. Section \ref{egpsec} introduces a new line
index for the CH G-band. The main features of the \hes{} database, as well as
the criteria for extracting the second subsample and the development of
saturation corrections for bright sources are presented in Section
\ref{database}. Section \ref{secanalysis} discusses the selection criteria
applied to the \hes{} database. The medium-resolution spectroscopic follow-up
observations and estimates of atmospheric parameters and carbon abundances are
described in Section \ref{followup}. Finally, our conclusions and perspectives
for future observational follow-up and further applications of the method are
presented in Section \ref{final}.

\section{EGP -- Additional Line Index for Carbon}
\label{egpsec}

One of the motivations for developing an additional line index for the CH G-band
is to increase the efficiency of the search for \cemp{} stars, and the very
metal-poor stars they are often associated with. The \gpe{} index from Paper I
has proven to be effective in searching for \cemp{} stars, but when used alone
it is subject to contamination from stars with spectra that have strong
H$\gamma$ lines (since this feature is included in the band covered by the
\gpe{} index) and from overlaps and plate artifacts on \hes{} plates.

Here we introduce a new line index for the CH G-band, as an alternative means of
measuring the contribution of this feature to the removal of flux in this
region, but without any dependence on continuum determination. This index is
calculated in a similar fashion to the definition proposed by \citet{smith1983}
and modified by \citet{morrison2003}. The 200\,{\AA} wide line band was chosen
following the same arguments found in Paper I for the \gpe{} index, which
measures the contrast between the G-band and a fitted continuum in the same
wavelength interval (see definition in Eqn. 1 of Paper I). The new \egp{}
index consists of a flux ratio, integrated in intervals defined for both the
line band and the red sideband, defined as:

\begin{equation}
{\rm EGP} = -2.5~{\rm log}\left[\frac{\int_{4200}^{4400}I_{\lambda_n}~d\lambda}{\int_{4425}^{4520}I_{\lambda_m}~d\lambda}\right]
\label{egpeq}
\end{equation}

\noindent where $I_{\lambda_n}$ and $I_{\lambda_m}$ are the measured flux (in counts)
for the line band and red sideband, respectively, and $d\lambda$ refers to the
spectral resolution, in \AA. The definition from \citet{morrison2003} uses
sidebands on both the red and blue sides of the line band. Due to the presence,
for cool stars (\teff~$\lesssim$~4500~K), of \cn{} bands on the blue side of the
line band (3883\,{\AA} and 4216\,{\AA}), the estimates for the integrated flux
can be strongly affected by these features, which could compromise the index.
Therefore, only the flux on the line band (200\,{\AA}{} wide; 4200-4400\,{\AA})
and the red sideband (4425-4520\,{\AA}) are used.

\begin{figure}[!ht]
\epsscale{1.00}
\plotone{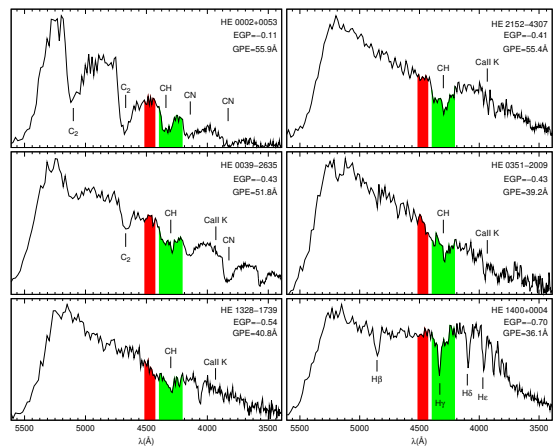}
\caption{Definition of the \egp{} index. The areas in
red and green represent the sideband and the line band, respectively, for HES
prism spectra of stars over a range of temperature and carbon enhancement.  Note that 
the abscissa runs from red to blue wavelengths.}
\label{indexegp}
\end{figure}

Figure \ref{indexegp} shows the wavelength intervals of the \egp{} index for the
prism spectra of six stars from the \hes{} plates. It is important to recall
that, in contrast to the \gpe{} index, \egp{} depends essentially on the depth
of the G-band and the flux difference when compared to the contiguous red region
of the spectra. It also possesses a smaller dynamical range (due to the
logarithmic ratio) when compared to the \gpe{} index. Nevertheless, this does
not present the same difficulties exhibited by the formerly employed GP and
GPHES indices \citep[definitions in][respectively]{beers1999,christlieb2008},
since the \egp{} does not use continuum estimates based on sideband linear
interpolations.

\begin{figure}[!ht]
\epsscale{1.00}
\plotone{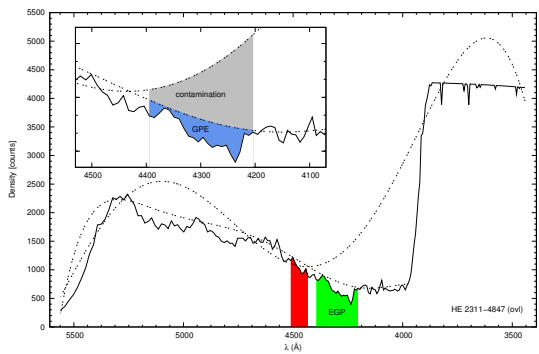}
\caption{Example of the {\emph{ovl}} class from the \hes, showing the 
definition of the \egp{} index. The dashed lines represent the continuum
calculated for the whole spectra and the continuum estimated without the
overlap. The expanded detail box shows a representation of the contamination of
the \gpe{} index by the overlapping spectra.}
\label{ovleps}
\end{figure}

From inspection of the center and bottom panels on the right side of Figure
\ref{indexegp}, one can see the advantage of introducing an auxiliary line index
for the selection of \cemp{} stars. While the two objects present \gpe{} values
that differ by no more than 10\%, the \egp{} difference reaches almost 50\%. In
those cases, the combination of the indices can be used as a filter for stars
with prominent H$\gamma$ lines. Furthermore, this set of indices can be used to
exclude spurious values of \gpe, caused by (among others) overlapping spectra.
Since both \gpe{} and \egp{} measure the same region using different
definitions, one should expect a reasonably strong correlation between the index
values calculated for a given star. So, any large deviations from this expected
behaviour can be interpreted as arising from possible contamination (see Section
\ref{visu} for more details). Figure \ref{ovleps} shows how this effect can
compromise the continuum calculation, and therefore the \gpe{} value. One can
see that the automated procedure fits the entire spectrum, and introduces a
contamination on the index (see expanded detail in Figure \ref{ovleps}). Thus,
with the help of the \egp{} value, which is less affected by the overlap, one
can exclude this type of object before going through the final visual
inspection.

\section{The HES Stellar Database}
\label{database}

Even though it was originally designed to detect bright (V $<$ 16.5) quasars
suitable for high-resolution spectroscopy, which drove the resolution
requirement up to the point it was also useful for stellar science
\citep{reimers1990,wisotzki2000}, the stellar data from the HES has been the
subject of a great number of studies over the past decade
\citep[][and others]{christlieb2001,christlieb2005,barklem2005,frebel2006,
aoki2007,christlieb2008,schorck2009,li2010,placco2010,kennedy2011}, 
and has provided candidates for photometric follow-up, as well as 
moderate- and high-resolution spectroscopy.

The \hes{} database presents a homogeneous, statistically well-understood sample
of stars which can be used to assess many interesting questions regarding the
origins of stellar populations and the formation of the Galaxy 
\citep[see][]{beers2005}. Both the spectral resolution (15\,{\AA} at 
Ca\,{\sc{ii}} K -- 3933\,{\AA}) and wavelength coverage (3200-5300\,{\AA}) of
the \hes{} spectra are suitable for searching for metal-poor stars by taking
advantage of what has been learned about the behavior of the CH G-band (4300\,
{\AA}) and Ca\,{\sc{ii}} K line at low metallicity.

\subsection{The Second HES Subsample}

The present study made use of two classes of objects extracted from the \hes{}
objective-prism plates: {\emph{stars}} and {\emph{bright}}. The {\emph{stars}}
are point-like sources showing no signs of saturation in their spectra and the
{\emph{bright}} sources are objects close to (and above) the saturation
threshold of the \hes{} plates. For each of these types, a different set of
restrictions was developed, including a more relaxed version of the method
described in \citet{christlieb2008}, and applied to the database in order to
reduce the number of spurious candidates for subsequent analysis.

\subsubsection{Source Type: Stars}
\label{subsubs}

The selection criteria for the {\emph{stars}} do not follow the same
prescriptions found in Paper I, in hopes of removing any known sample-related
biases, specially regarding the photometric quality and apparent magnitudes of
the objects. The restriction on source type removes bright and extended sources
from the \hes{}. These extended sources can contaminate the selected sample with
galaxy candidates, and the {\emph{bright}} sources have a distinct selection
criteria, described later in this section.

The \bvz{} colors were retrieved from the \hes{} database, using the improved
calibration described in \citet{christlieb2008}. A selection was then made in
the color range 0.30~$\leq$\bvz$\leq$~1.00, in order to avoid stars outside the
optimal color range for the atmospheric parameters calculation (see Section
\ref{ssppsec}). 

The J and K magnitudes were taken from the 2MASS All-Sky Data Release
\citep[Two Micron All Sky Survey; ][]{2mass}, and were used only for the objects
labeled with the photometric quality flags ``A'', ``B'' or ``C''. The
de-reddened \jk{} colors were calculated based on the \citet{schlegel1998}
prescription. The selected objects lie in the color range
0.20~$\leq$\jk$\leq$~0.75. Finally, restrictions were placed on the average
signal-to-noise ratio in the calcium line region (SN$_{\rm CaHK}>$~5), 
as well as on the KP index, which was chosen to be greater than 
$-1.0\cdot \sigma_{\mathrm{KP}}$, where $\sigma_{\mathrm{KP}}$ is 
the detection limit for the Ca\,{\sc{ii}}~K line \citep[see Figure 6
of][]{christlieb2008}. The latter criterion aims at rejecting spectra that
exhibit a Ca\,{\sc{ii}}~K line in emission, and for which negative KP indices are
measured.

The \hes{} {\emph{stars}} fulfilling all of the above restrictions and contained
inside both color intervals were then passed through the selection procedure in
the KP/\bvz{} and KP/\jk{} planes described in detail by \citet{christlieb2008}.
However, the selection adopted in this work sets a more relaxed criterion for
the metallicity cut, setting it at \metal\ =$-$2.0, rather than the more
conservative limit of \metal\ =$-$2.5 employed by \citet{christlieb2008}. The
objects selected were required to be found within the cut regions for at least
one of the KP/color planes in order to be recognized as a metal-poor candidate.
The final subset of the source type {\emph{stars}} contains 62,311 candidates.

\subsubsection{Source Type: Bright}
\label{subsubb}

The criteria for selecting {\emph {bright}} sources is very similar to the one
applied to the {\emph {stars}}. The only difference is the exclusion of the
KP/\bvz{} restriction, due to the fact that saturation effects play a major role
on the accuracy of color determinations for the brightest stars from the \hes{}
\citep{frebel2006}. Furthermore, since the absolute number of {\emph {bright}}
sources in the \hes{} is on the order of 1/6th the number of {\emph {stars}},
there was no need to introduce additional criteria to filter down the number of
{\emph {bright}} objects for visual inspection. The final subset of the source
type {\emph{bright}} contains 18,532 candidates.

One of the main goals of the present analysis is to apply a single set of
contraints for both {\emph {stars}} sources and for {\emph {bright}} . However,
as mentioned by previous studies \citep[e.g.][]{frebel2006}, the {\emph
{bright}} objects from the \hes{} suffer saturation effects, which have to be
dealt with before proceding with the analysis of the entire candidate list.

\subsection{Saturation Corrections for Bright Sources}
\label{satsec}

Saturation can effect most astronomical data, at least when photons are
collected by material (photographic plates or CCDs) with a finite capacity for
measuring light. For photographic plates the problem is actually more severe,
due to non-linearities that begin to develop in the response curves even before
the saturation level is reached. Hence, line indices that rely on comparisons of
the relative densities of regions of spectrum on photographic plates will vary
somewhat, even when stars of the same intrinsic compositions, temperatures, and
surface gravities, but differing apparent magnitudes, are measured.

\begin{figure}[!ht]
\epsscale{1.00}
\plotone{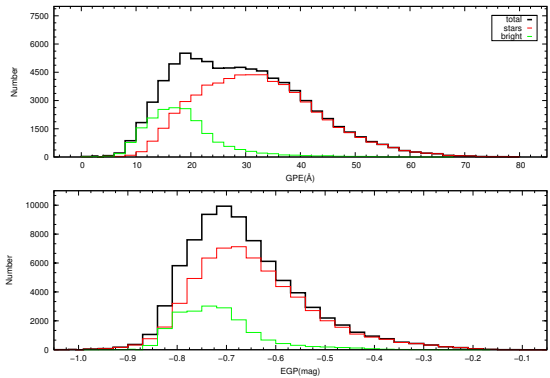}
\caption{Distribution of the CH G-band line indices for the second \hes{} subsample,
divided by source type as shown in the legend.}
\label{histold}
\end{figure}

Figure \ref{histold} shows the distribution of the line indices for the 80,843
objects of the second \hes{} subsample, divided by source type. It is clear that
both \gpe{} and \egp{}, as calculated for the candidates, do not share the same
behaviour for the {\emph {bright}} sources as seen for the {\emph {stars}}. One
might suppose mistakenly that the {\emph {bright}} sources have intrinsically
less carbon than the {\emph {stars}}, which seems very unlikely. Rather, the
histograms are misleading because the {\emph {bright} sources suffer much more
than the {\emph {stars}} from the saturation effects mentioned above, which have
compromised their index calculations. 

To compensate for this saturation issue for the {\emph {bright}} sources, a
numerical correction was developed, taking into account not only individual
effects, but also trends for groups of objects. This correction was made by
means of the {\emph{spcmag}} variable from the \hes{}, which is an internal
magnitude based on the integrated photographic density in the B$_{\rm J}$ band.
Determination of the particular value of B$_{\rm J}$ at which saturation occurs
will vary, depending on the sensitivity of the individual \hes{} plate, seeing
conditions under which it was taken, and many other factors. As a result, this
magnitude cannot be used to set the corrections for the line indices. However,
{\emph{spcmag}} is a good global indicator for the level of saturation; i.e.,
this indicator should be valid for all plates, since it measures the
photographic density associated with the stars on each plate.

The distributions of B$_{\rm J}$ magnitude and {\emph{spcmag}}, for both
saturated and non-saturated sources, are shown in Figure \ref{histspcbj}. As can
be appreciated from inspection, the two source types do not even share the same
range of values. While the {\emph {stars}} present a similar distribution as
previous searches for metal-poor candidates \citep[see Figure 10
of][]{christlieb2008}, the {\emph {bright}} sources appear in the same magnitude
interval (9 $< {\rm B_J} <$ 14) as the bright objects studied by
\citet{frebel2006}. Thus, the hypothesis is put forward that, for a given
interval of {\emph{spcmag}}, the {\emph{bright}} sources will present the same
saturation level for the line indices. That is, by looking at the index
distribution of each {\emph{spcmag}} interval of saturated and non-saturated
sources, it is possible to estimate the amount of this saturation that
translates into the index values. Then, to correct this effect, one should shift
the distribution of line indices for each interval and match the distribution of
the non-saturated sources.

\begin{figure}[!ht]
\epsscale{1.00}
\plotone{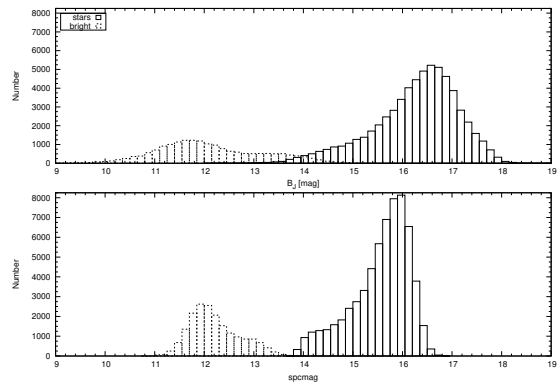}
\caption{Distribution of B$_{\rm J}$ (BHES) and 
{\emph{spcmag}} for the second \hes{} subsample, divided by source type as shown
in the legend.}
\label{histspcbj}
\end{figure}

The sample containing the {\emph {bright}} sources (18,532 objects) was divided
into nine parts according to the {\emph{spcmag}} values. The normalized
distributions for each part were compared with the normalized distribution of
the non-saturated sources (containing all the 62,311 objects classified as
{\emph{stars}}). Figures \ref{chigpe} and \ref{chiegp} show, respectively, the
distributions of the \gpe{} and \egp{} indices for the individual parts. The
black histograms represent the distribution for the non-saturated sources, and
the ones in green are the index distributions for each of the nine divisions in
{\emph{spcmag}}. The histograms in red represent the corrected distributions.
For each individual distribution, GPE$_{\rm I}$, EGP$_{\rm I}$, GPE$_{\rm F}$
and EGP$_{\rm F}$ are, respectively, the maximum values of the saturated (I) and
non-saturated (F) distributions of the \gpe{} and \egp{} indices. Thus, the
saturation corrections for each division in {\emph{spcmag}} are quantified by
the required horizontal shift of each distribution, so that the saturated and
non-saturated distributions match.

\begin{figure}[!ht]
\epsscale{1.0}
\plotone{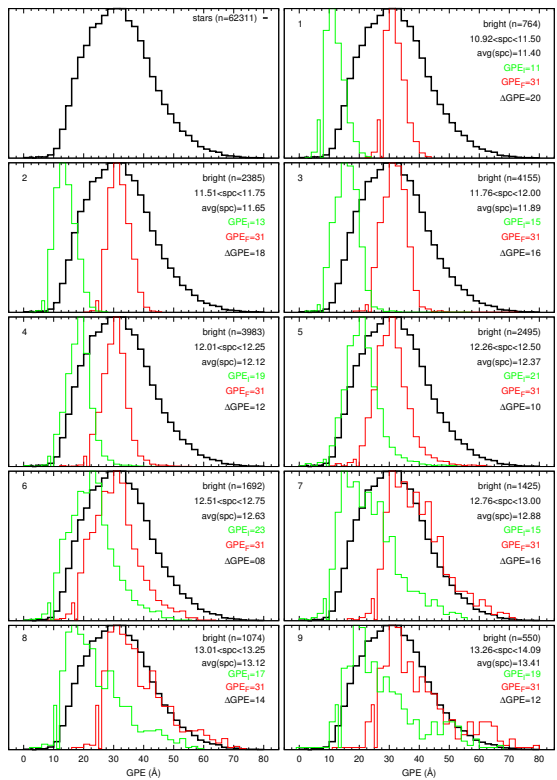}
\caption{Saturation corrections for \gpe, shown over the range of
{\emph{spcmag}}. The black histograms represent the distribution for the
non-saturated sources, and the ones in green are the index distributions in
{\emph{spcmag}}. The number of objects, magnitude range, and average values are
shown on the upper right of the panels. The histograms in red represent the
corrected distributions. GPE$_{\rm I}$, EGP$_{\rm I}$, GPE$_{\rm F}$, and
EGP$_{\rm F}$ are, respectively, the maximum values of the saturated (I) and
non-saturated (F) distributions of the \gpe{} and \egp{} indices. Thus, the
saturation corrections for each division in {\emph{spcmag}} are quantified by
the horizontal shift of each distribution, listed in each panel as $\Delta$GPE,
required to make the saturated and non-saturated distributions match.}
\label{chigpe}
\end{figure}

\begin{figure}[!ht]
\epsscale{1.0}
\plotone{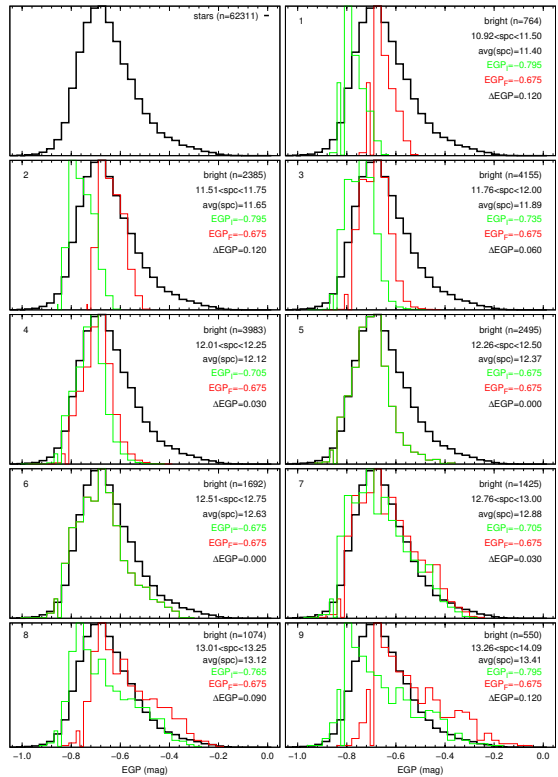}
\caption{Saturation corrections for \egp, shown over the range of
{\emph{spcmag}}.  The approach is identical to that described for Figure \ref{chigpe}, but
for the \egp{} index.}
\label{chiegp}
\end{figure}

It is interesting to note that the shifts in the \egp{} distributions are
systematically smaller than the ones for the \gpe, and can be as low as zero for
12.26~$<spcmag({\rm mag})<$~12.70. This can be understood due to the differences
in the calculations of the line indices, because the saturation has a greater
effect on the continuum calculations than on the flux ratio. Besides that, the
shifts tend to decrease with increasing {\emph{spcmag}}, up to $\sim$12.75~mag,
when the values start to increase. From inspection of Figures
\ref{chigpe} and \ref{chiegp}, one can conclude that this sudden increase (for
{\emph{spcmag}}$>$~12.75~mag) is mainly because of the shape of the
distributions. Given that those distributions contain fainter objects,
associated with lower signal-to-noise spectra, the calculated index values are
affected by the quality of the spectra and become more scattered. Hence, we chose
to set aside the shifts determined for the last three distributions of Figures
\ref{chigpe} and \ref{chiegp}. For those (panels 7, 8 and 9), the shifts used
were the ones associated with the distribution shown in panel 6.

Once determined, the shifts associated with each distribution of {\emph{spcmag}}
can be used to calculate the correction functions for \gpe{} and \egp{}, by
making polynomial fits to the [~avg(spc)~,~$\Delta$GPE~] and [~avg(spc)~,
~$\Delta$EGP~] data pairs, as shown in Figure \ref{compcor}. There were a number
of attempts to fit polynomials of degree 1-5 to the data available , and the
lower values of the asymptotic standard errors for the final set of parameters
are associated with a 4th degree polynomial (${\rm F_{GPE}}$ and ${\rm
F_{EGP}}$). However, even this fit is not good for {\emph{spcmag}}~$<$~11.40,
where there is a sudden decrease of the data, and {\emph{spcmag}}~$>$~12.63,
that shows an unexpected increase of the shift values. Therefore, the
corrections applied to the objects with {\emph{spcmag}}~$<$~11.40 are those
associated with the smaller values of avg(spc): 20~\AA{} for \gpe{} and 0.12~mag
for \egp{}. For the objects with {\emph{spcmag}}~$>$~12.63, 8\AA{} is used for
\gpe{} and for \egp{} there are no corrections made. The final saturation
correction criteria for the \gpe{} and \egp{} indices are:

\begin{figure}[!ht]
\epsscale{1.00}
\plotone{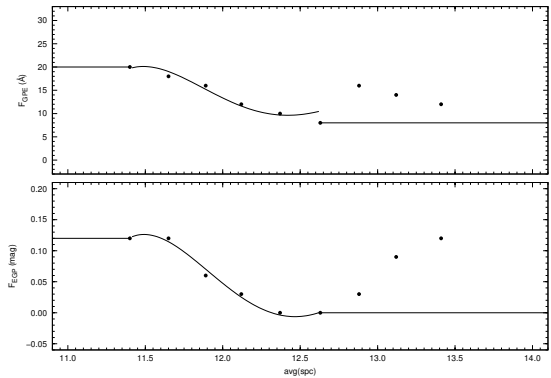}
\caption{Proposed correction function
for the line indices, as a function of the average value of {\emph{spcmag}} for
each part of the distribution.  Note that the last three points in each panel
are not used (see text for explanation).}
\label{compcor}
\end{figure}

$$
\Delta{\rm GPE(\AA)}= \left\{ \begin{array}{rr}
20.0, & spcmag<11.40 \\
{\rm F_{GPE}}, & 11.40\leq spcmag \leq12.63 \\
8.0, & spcmag>12.63
 \end{array} \right.
 \vspace{0.3cm}
$$
$$
\Delta{\rm EGP(mag)}= \left\{ \begin{array}{rr}
0.12, & spcmag<11.40 \\
{\rm F_{EGP}}, & 11.40\leq spcmag \leq12.63 \\
0.00, & spcmag>12.63
 \end{array} \right.
\vspace{0.3cm}
$$

\noindent where F$_{\rm GPE}$ and F$_{\rm EGP}$ are given by:

\begin{eqnarray}
{\rm F_{GPE}}= -371276 + 120767 \cdot {\rm [avg(spc)]} \nonumber \\ 
-14707.1 \cdot {\rm [avg(spc)]}^2 + 794.768 \cdot {\rm [avg(spc)]}^3 \nonumber \\ 
-16.081 \cdot {\rm [avg(spc)]}^4 \\
{\rm F_{EGP}}= -3527.320 + 1139.050 \cdot {\rm [avg(spc)]} \nonumber \\ 
-137.652 \cdot {\rm [avg(spc)]}^2 + 7.378 \cdot {\rm [avg(spc)]}^3 \nonumber \\
 -0.148 \cdot {\rm [avg(spc)]}^4
\end{eqnarray}

With the saturation corrections applied to the {\emph {bright}} sources, it was
possible to rebuild the line indices distributions for the second subsample
objects. Results are shown in Figure \ref{histnew}. In comparison to Figure
\ref{histold}, one can note the change in the shape of the corrected
distributions for the {\emph {bright}} sources, which is a direct effect of the dependency
with {\emph{spcmag}}. Before the saturation corrections, the average values of
the \gpe{} and \egp{} indices for the {\emph {bright}} sources were, respectively, 18.7\,
{\AA} and $-$0.72~mag. After the corrections, the values are 31.3\,{\AA} (32.3\,
{\AA} for the {\emph{stars}}) and $-$0.68~mag ($-$0.65~mag for the
{\emph{stars}}), showing that the line index distributions for the {\emph
{bright}} sources are in agreement with the non-saturated {\emph {stars}}.

\begin{figure}[!ht]
\epsscale{1.00}
\plotone{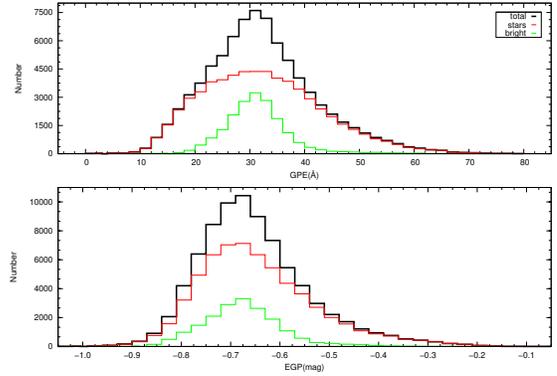}
\caption{Distribution of the saturation-corrected line indices, divided by source type
as shown in the legend.}
\label{histnew}
\end{figure}

\section{Selection of Metal-Poor Candidates}
\label{secanalysis}

The saturation corrections described in Section \ref{satsec} bring both source
types of the second subsample to a common scale, so it is possible to apply a
single set of restrictions for \gpe{} and \egp{} to search for metal-poor stars
based on carbon enrichment. This procedure is similar to the one adopted by
\citet{christlieb2001}, who used a pair of carbon molecular indices (for \cn{}
and \ctwo) to find carbon-rich stars based on low-resolution spectra. The
advantages of working with two indices defined in the same region of the
spectra, but with different approaches in representing it, lies in the
opportunity to identify likely spurious values, as discussed above.

\subsection{Selection Criteria} 

The last step of the selection criteria has the goal of restricting the
parameter space for the corrected line indices and provide a suitable subsample,
both in absolute number and fraction of \cemp{} candidates, for the visual
inspection. The second subsample contains 41,454 objects with
\gpe~$>$~31\,{\AA} and 42,823 objects with \egp~$>$~$-$0.675, with 31,889
satisfying both conditions. It is interesting to note that for the sample in Paper
I, only $\sim$7$\%$ of the stars were selected after the restriction in \gpe.
For the second subsample, this fraction is $\sim$51$\%$, even after the
selection in the KP/color plane, which allows a more narrow index range to
search for metal-poor stars.

In order to decrease the number of objects for visual inspection, another cut in
\jk{} was made, since the subsample has a great number of cool stars and we did
not want to impose a more restrict criteria for the metallicity in the KP/color
plane. as stated above. Besides that, for stars with \jk~$>$~0.7, the accuracy 
of atmospheric parameters
are limited in medium-resolution spectroscopy, mainly due to the low intensity
of the Balmer lines used as auxilliary temperature indicators
\citep{schorck2009}, as well as the increasing saturation of the Ca\,{\sc{ii}} K line.

\begin{figure}[!ht]
\epsscale{1.00}
\plotone{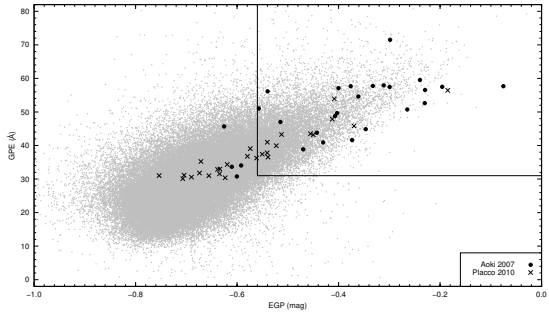}
\caption{Behavior of the line indices for the \hes{} second subsample, 
along with the \cemp{} stars from the high-resolution spectroscopic follow-up
of \citet{aoki2007} , and the medium-resolution follow-up reported in Paper I.
The solid lines represent the restrictions applied to the indices.}
\label{criterio2}
\end{figure}

Figure \ref{criterio2} shows the behavior of the \gpe{} and \egp{} indices for
the second \hes{} subsample stars with \jk~$<$~0.7, along with the values for
the low-resolution spectra of the \cemp{} stars (\metal$<-$1.0 and
\cfe$>+$1.0) from \citet{aoki2007} and Paper I. The solid lines show
the restrictions for the line indices. It is possible to see that some of the
confirmed \cemp{} were left out of the selection window. This was a compromise
between the number of candidates for visual inspection and the possible fraction
of \cemp{} stars to be found within it.

The final restrictions applied to the line indices and color for the second
subsample are: (i) \gpe~$>$~31\,{\AA}; (ii) \egp~$>$~$-$0.56 mag and; (iii)
\jk~$<$~0.7 mag. This set of constraints yielded 10,314 objects that were
subjected to visual inspection, as described in the following subsection.

\subsection{Visual Inspection}
\label{visu}

The visual inspection of the selected objects of the second subsample was
carried out following the same methodology presented in Paper I, sorting the
candidates based on spectral absorption features on the \hes{} 1D spectra and
digitized plates. The overall appearance of the object on the DSS direct plates
was also inspected, to track down overlaps and plate artifacts. Besides that, no
changes were made in the procedure due to the inclusion of {\emph {bright}}
sources, and the objects were divided into the same classes used in
\citet{christlieb2008} and Paper I.

\subsubsection{Main Features and Results}

The results of the inspection, and a brief description of the classes, are
presented in Table \ref{canclass}. One can notice that the second subsample is
dominated by stars with strong Ca\,{\sc{ii}}~K lines ({\emph{mpcc}}), which is a
direct reflection of the cumulative metallicity distribution function. There is
also a great number of low-S/N spectra in the sample. This is due to the fact
that there was no restriction made on the apparent magnitudes of the objects, in
order to sample stars either more distant or less luminous, and also the outer
regions of the Galaxy, including the dual halo components \citep{carollo2007,
carollo2010, carollo2011}. Moreover, the classes {\emph{mpca}}, {\emph{mpcb}},
and {\emph{fhlc}} received a greater fraction of candidates than in Paper I,
which means that the number of stars with \metal~$<-$2.0 should increase when
compared to the previous effort.

Another important point that arises from the quantities in Table \ref{canclass}
is the fact that there are only 3 objects, out of the over 10,000 inspected,
presenting strong hydrogen lines (class {\emph{habs}}). This is due, in part,
to the \jk~$>$~0.2 restriction, but mainly because of the combination of the
\gpe{} and \egp{}. As shown in Section \ref{egpsec}, the relation of the line
indices for objects with prominent hydrogen lines tends to be different then the
one for the \cemp{} candidates, so this means that the restrictions imposed
successfully filtered out these contaminations.

Figure \ref{plot2ndA} shows the distribution of the line indices for the
inspected stars, excluding classes {\emph{nois}}, {\emph{ovl}}, and {\emph{art}}.
The remaining classes of Table \ref{canclass} are evenly distributed about a
linear relation between the line indices, with exception of {\emph{mpca}} and
{\emph{mpcb}}, which are concentrated in the lower left region of the plot,
along with the three objects associated with strong hydrogen lines. The stars
with strong Ca\,{\sc{ii}}~K lines ({\emph{mpcc}}) exhibit a wide range of index
values. So, one can conclude that these objects would present a variety of
\cfe{} values, and also \metal{} values, since even cool giant stars with
\metal\ $\sim-$2.0 exhibit strong Ca lines \citep{schorck2009}. In addition, when
comparing Figure \ref{plot2ndA} with the same region in Figure \ref{criterio2},
one can note a clear trend between the line indices, and the objects located far
from this trend (i.e., \gpe~$<$~40\,{\AA} and \egp~$>$~0.3 mag) have spectra
associated with the classes excluded from the plot, with spurious values for the
indices.

\begin{center}
\begin{deluxetable}{ccc}[!ht]
\tablecaption{Visual Inspection Classification for the Selected Candidates \label{canclass}}
\tablehead{Tag & Description & Candidates}
\tablewidth{0pt}
\startdata
mpca & Absent Ca\,{\sc{ii}} K line        &   55  \\
mpcb & Weak Ca\,{\sc{ii}} K line          &  452  \\
mpcc & Strong Ca\,{\sc{ii}} K line        & 5184  \\
unid & Ca\,{\sc{ii}} K line not found     &  224  \\
fhlc & Faint high latitude carbon stars   &  317  \\
habs & Strong absorption H lines          &    3  \\
hbab & Horizontal-branch/{}$A$ type star  &    0  \\
nois & Low signal-to-noise ratio          & 2320  \\
ovl  & Overlapping spectra                & 1038  \\
art  & Artifacts on photographic plates   &  721  \\
\enddata
\end{deluxetable}
\end{center}

\begin{figure}[!ht]
\epsscale{1.00}
\plotone{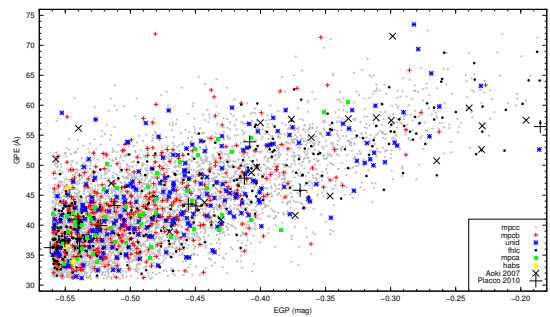}
\caption{Results of the visual inspection of the candidates,
according to the classes described in Table \ref{canclass}.}
\label{plot2ndA}
\end{figure}

\subsubsection{Comparison with Paper I}

We also wish to test the efficiency of our new search procedure, using the
\gpe{} and the \egp{} indices, by comparing the results of the visual 
inspection with the one made in Paper I. This can be done by calculating the
effective fractions of the various types of objects identified, excluding the
classes {\emph{art}}, {\emph{nois}} and {\emph{ovl}}. Results are shown in Table
\ref{canclasscomp}. Some interesting aspects of the visual inspection can arise
from the analysis of the fractions presented: (i) the ``clean'' samples present
a similar number of candidates, but the one associated with this work does not
have magnitude restrictions. This removes the brightness and/or distance related
bias introduced in Paper I due to observational limitations; (ii) the fractions
associated with the {\emph{mpcc}} class for both subsamples are similar,
suggesting that those stars are evenly distributed in the parameter space
associated with the visual inspections; and (iii) for the {\emph{mpca}} and
{\emph{fhlc}} classes, the fractions increased by an order of magnitude in
comparison to the Paper I sample. These candidates form a set of stars with high
probabilities of occurance of EMP (\metal~$<-$~3.0) carbon-rich stars; (iv) the
objects with intense hydrogen lines ({\emph{habs}} and {\emph{hbab}}),
considered a contamination on the first subsample, were (with the exception of
three objects) removed from the second subsample by the method based on the
combination of line indices and by the restriction imposed in \jk{} color; and (v)
the objects associated with the {\emph{mpcb}} and {\emph{unid}} classes
presented an increase of about $\sim$30\% in their effective fractions.

\begin{center}
\begin{deluxetable}{c|cc|cc}
\tablecaption{Comparison Between Visual Inspections \label{canclasscomp}}
\tablehead{ & \multicolumn{2}{c|}{Paper I sample} & \multicolumn{2}{c}{This work} \\ Class & $n$ & fraction & $n$ & fraction}
\tablewidth{0pt}
\startdata
mpca & 4       & 0.07$\%$   & 55   & 0.88$\%$ \\
mpcb & 293   & 5.32$\%$   & 452  & 7.25$\%$ \\
mpcc & 4711 & 85.58$\%$ & 5184 & 83.14$\%$ \\
fhlc   & 31     & 0.56$\%$   & 317   & 5.08$\%$ \\
unid  & 155   & 2.81$\%$    & 224 & 3.60$\%$ \\
hbab & 235   & 4.28$\%$   & 0     & 0.00$\%$ \\
habs & 76     & 1.38$\%$   & 3     & 0.05$\%$ \\
\hline
Total & 5505 & 100$\%$ & 6235 & 100$\%$ \\
\enddata
\end{deluxetable}
\end{center}

Finally, a search in the literature was made in order to exclude objects already
observed or classified by other studies. Among the 6,235 stars shown in Figure
\ref{plot2ndA}, 5,288 do not appear in any study to date. A table with relevant
information on these stars is available electronically, and its parameters are
listed in Appendix \ref{aplist}.

\section{Follow-up Observations of Selected Candidates}
\label{followup}

To validate the analysis of our selected candidates, especially for the
saturation-corrected {\it bright} sources, we have obtained medium-resolution
spectra for a limited number of metal-poor candidates with the SOAR 4.1m and
Gemini 8m telescopes. After gathering and reducing the data, we obtained
estimates of the stellar atmospheric parameters using the n-SSPP, a modified
version of the SEGUE Stellar Parameter Pipeline \citep[SSPP - see][for a
detailed description of the procedures used]{lee2008a,lee2008b,allende2008,
smolin2011,lee2011}. The carbon abundances (\cfe), were obtained using spectral
synthesis. Further details are provided below.

\subsection{Spectroscopic Observations and Stellar Data}

The observed sample consists of data collected with two different
telescope/spectrograph combinations: SOAR/Goodman and Gemini/GMOS. The addition
of Gemini observations were of great interest, specifically to increase the
magnitude limit reached by our program to B$\sim$16.0, with exposure times no
longer than 30 minutes. Accordingly, the brighter sources were observed
preferentially with SOAR.

The 58 star sample from SOAR consists of data observed in the 2009B, 2010B and 2011A
semesters. All the observations were carried out with the Goodman Spectrograph,
using the 600~l~mm$^{\rm{-1}}$ grating in the blue setting with a 1$\farcs$03 slit,
covering the wavelength range 3550-5500\,{\AA}. This combination yielded a
resolving power of $R\sim 1500$ and sufficient signal-to-noise ratios
(S/N$\sim 40$ per pixel at 4300 \AA). GMOS data for the 59 stars from Gemini North and
South were gathered in 2010A and 2011A, observed in Bands 3 and 4. The setup was
similar to the one used in SOAR, with a 600~l~mm$^{\rm{-1}}$ grating in the blue
setting (G5323 for GMOS South and G5307 for GMOS North) and a 1$\farcs$00 slit.
The resolving power ($R\sim 1700$) and signal-to-noise ratios (S/N$\sim 40$ per
pixel at 4300 \AA) of these spectra are also suitable to determine atmospheric
parameters and carbon abundances.

For both SOAR and Gemini data, the calibration frames included HgAr and Cu arc
lamp exposures (taken following each science observation), biases frames, and quartz
flats. All tasks related to calibration and spectral extraction were performed
using standard IRAF packages. Table \ref{candlist} lists the equatorial
coordinates, BHES magnitude, \jk{}, \gpe{}, \egp{}, telescope and target 
classification (according to Table \ref{canclass}) for the 117 observed candidates.

\subsection{Atmospheric Parameters, \cfe{} and \cemp{} Candidates}
\label{ssppsec}

The atmospheric parameters and carbon abundances were determined with the same
procedures used in Paper I. Given the improvements of the n-SSPP over the last
two years, we decided to reprocess the data for the first sample, in order to
produce a more homogeneous set of parameters. 

Figure \ref{goodman132_comp} shows the residual distribution of \metal{},
\teff{}, and \logg{} as a function of the latest estimates. The residuals for the
metallicity spread between $-$0.5 and 0.5 dex for \metal\ $<-$1.0, and there is a
strong positive trend for the new calculations on the metal-rich end. These new
values of \metal{} also reflect on the \cfe, since the estimated values of
\abund{C}{H} did not change. For the temperature residuals there are some values outside
the [$-$500~K:$+$500~K] range, and for \logg{} there is an underlying positive
linear trend for increasing values. These changes are mainly due to new
calibrations implemented on the code, and extensive testing was performed to
assure the quality of the new determinations.

\begin{figure}[!ht]
\epsscale{1.00}
\plotone{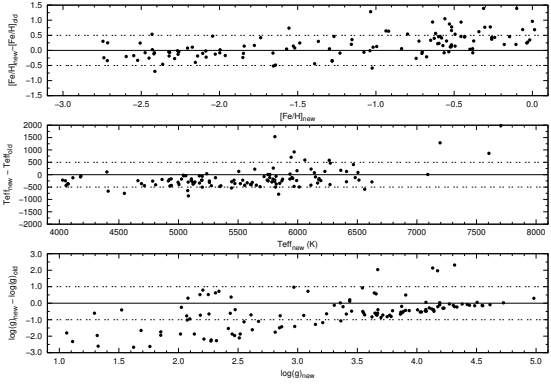}
\caption{Residual distribution of the atmospheric 
parameters for the Paper I sample, reprocessed with the latest version of the n-SSPP.
The trends represent recent improvements made to the n-SSPP.}
\label{goodman132_comp}
\end{figure}

The atmospheric parameter estimates for the observed metal-poor candidates
presented in Table \ref{candlist} were also obtained via the n-SSPP; results are
listed in Table \ref{atmpar}. The last column refers to the carbon-to-iron
abundance ratios (\cfe; ``carbonicity''), obtained with the same procedure
described in Paper I, using spectral synthesis in the G-band region. Updated
results for the sample from Paper I are also shown in the table. Typical errors for
the atmospheric parameters are 0.3 dex for \metal, 300~K for \teff{} and 0.4 for \logg. 
The errors associated with the carbonicity are, on average, 0.2 dex.

Figure \ref{carbon} shows the behavior of carbonicity as a function of
metallicity for the stars observed in this work, as well as from the
medium-resolution data of Paper I, high-resolution spectroscopic data from
\citet{aoki2007} and \citet{caffau2011}, 
and for stars with \metal$<-2.5$ retrieved from the SAGA database 
\citep[Stellar Abundances for Galactic Archeology -][]{saga2008}. 
The left and lower panels show the distributions of \metal{}
and \cfe{} for the stars of this work and Paper I. It can be seen that
the new selection criteria successfully sampled the \metal$<-$3.0 region, with
essentially all stars in this range presenting \cfe\ $>+$1.0, including two stars
with remarkably high carbonicity (\cfe\ $ \geq +$2.90). We also reproduce the strong
signature of increasing \cfe{} with declining \metal{} reported by previous
work. The increasing scatter in \cfe{} at lower metallicities cannot be
accounted for by uncertainties in the carbonicity determinations, rather it is
likely due to the presence of multiple nucleosynthesis paths that produce carbon
at low metallicity \citep{cescutti2010}. The features presented by this group of
stars are also consistent with the latest studies regarding the relationship
between the carbonicity and metallicity with the structural components of the
Milky Way. According to \citet{carollo2011}, the CEMP stars with metallicity
\metal~$< -2.5$ have a high probability of belonging to the outer-halo component
of the Galaxy. Future high-resolution spectroscopic follow-up of these stars can
help shed light on the matter.

\begin{figure}[!ht]
\epsscale{1.00}
\plotone{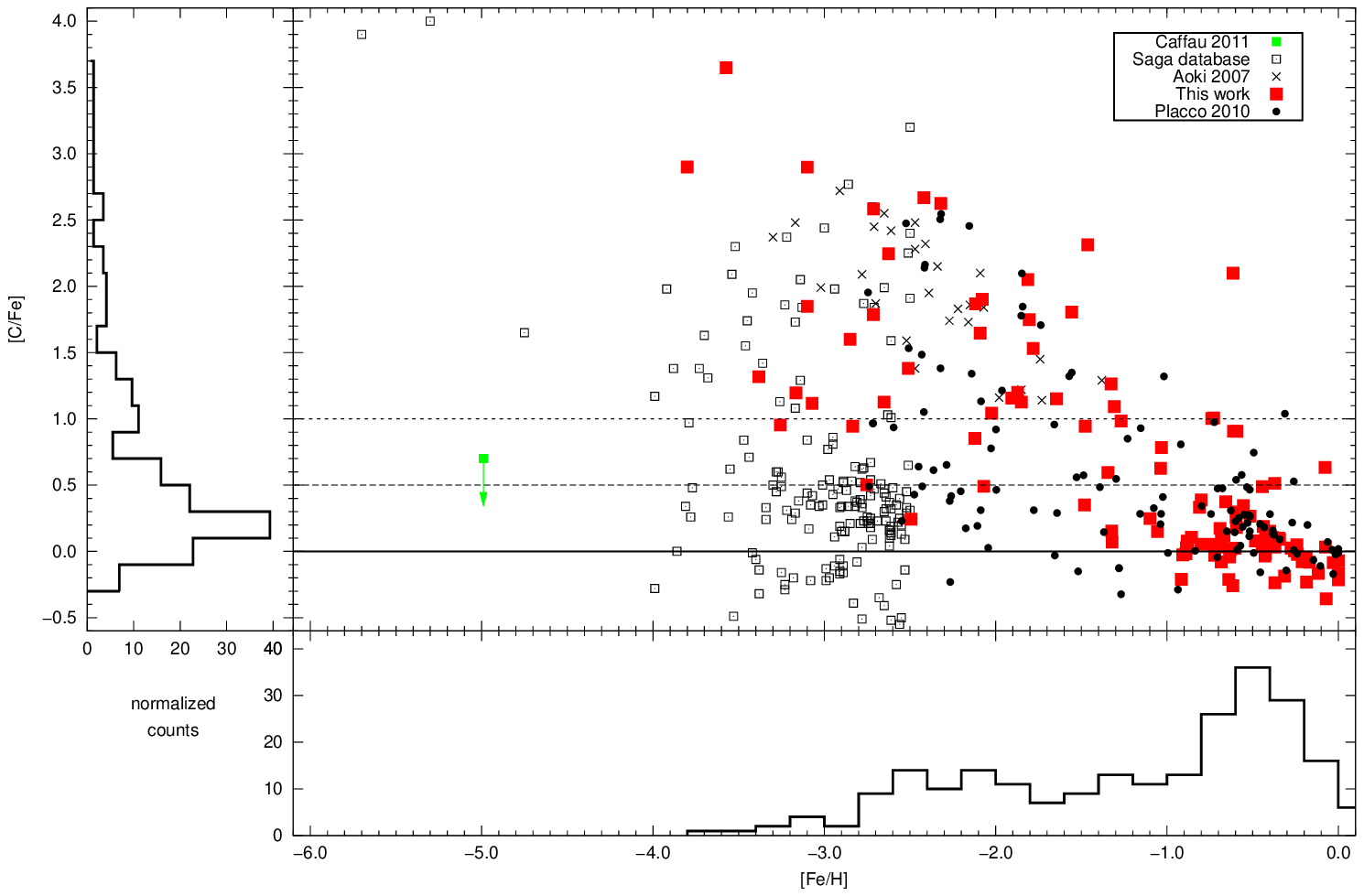}
\caption{Carbonicity, \cfe, as a function of the metallicity, \metal,
for the stars observed in this work, compared with data from the high-resolution
spectroscopic follow-up of \citet{aoki2007} and \citet{caffau2011}, from the SAGA database
and from the medium-resolution follow-up of Paper I.  The solid line is
the \cfe{} = 0.0 level, the long dashed line is the \cfe{} = $+$0.5 level, and
the short dashed line is the \cfe{} = $+$1.0 level. The left and bottom panels
show the marginalized distributions for each quantity, related to the data from 
this work and Paper I.}
\label{carbon}
\end{figure}

Even with the observation of a number of solar-metallicity stars, it is clear
from inspection of Figure \ref{carbon} that the new selection criteria is
successful in the discovery of VMP and EMP stars. This is mainly due to the
\gpe{} and \egp{} cuts shown in Figure \ref{criterio2}. By taking the index
values for the stars from SAGA database in Figure \ref{carbon} with
low-resolution counterparts in the \hes{} database, it can be seen that
roughly 70\% of the confirmed \cemp{} stars (\metal\ $<-1.0$ and \cfe\ $>+1.0$) fall
into the selection area (\gpe $>$ 31\,{\AA} and \egp\ $>$ -0.56 mag).

Figure \ref{specomp} illustrates two of the most interesting new discoveries of
our new approach, comparing the low-resolution \hes{} spectra (subject to visual
inspection) with data from SOAR and Gemini medium-resolution spectroscopy.
HE~1046$-$1352 (left panels) was selected after combining the two line indices
and separating the $mpca$ candidates by visual inspection. The right panels of
Figure \ref{specomp} show HE~1937$-$6314. This object was selected after
inclusion of the {\emph {bright}} sources (its BHES apparent magnitude is 13.6)
and subsequent saturation corrections of the line indices, and it is not the
subject of any previous study. Stars of this sort are of special interest, since
their apparent magnitudes permit shorter exposures for high-resolution
spectroscopic studies.

\begin{figure}[!ht]
\epsscale{1.00}
\plotone{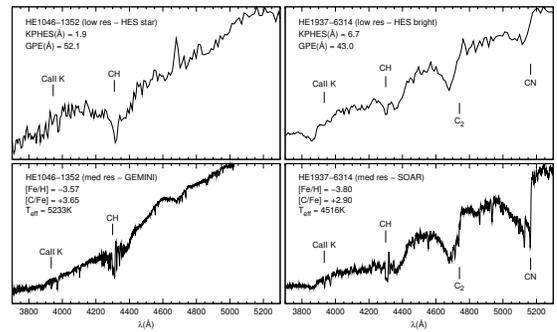}
\caption{Comparison between low-resolution \hes{} spectra and the medium-resolution 
spectra acquired with SOAR and Gemini. The main features assessed by visual
inspection are labeled in the upper panels.  HE~1046$-$1352 has BHES = 15.3, while
HE~1937$-$6314 has BHES = 13.6.}
\label{specomp}
\end{figure}

\section{Conclusions}
\label{final}

One of the goals of this work was to discover additional \cemp{} stars (which
are in turn often of very low metallicity) still hiding in the scanned
photographic plates and low-resolution spectra of the
\hes{}. To accomplish this, we extended the approach taken in Paper I to
include bright sources in the selection criteria applied to the database.

Line indices and corrections for the bright sources were developed specifically
for the \hes{} data, in order to compensate for the saturation effects presented
by these spectra. These will be included in the full stellar database, to help
in future searches for bright \cemp{} and metal-poor stars. Moreover, the
introduction of this new set of carbon line indices for low-resolution
spectroscopy can be easily adapted and implemented to help pre-process and
select interesting targets for follow-up spectroscopy in the massive amount of
data coming from the next generation of large surveys. The use of an auxilliary
index to search for \cemp{} candidates successfully filtered out some unwanted
objects, and increased the relative number of the most interesting classes of
metal-poor stars (see Table \ref{canclasscomp}). 

Medium-resolution spectroscopy with the SOAR and Gemini telescopes revealed an
improvement in our ability to isolated the targets of greatest interest relative
to Paper I. Comparing the stars of this work lying in the same magnitude range
as the targets from Paper I ($13.9 \le {\rm BHES} \le 15.1$), the fractions of
low-metallicity stars found increased from 53\% to 58\% for \metal~$< -1.0$,
from 27\% to 32\% for \metal~$< -2.0$ and from 7\% to 21\% for \metal~$< -2.5$,
an improvement by a factor of 3 in this last metallicity range. Our observations
of the newly identified sample of low-metallicity candidates has contributed
eight new stars with \metal~$< -3.0$.   

Regarding the carbonicity of our targets, over the entire magnitude range
observed by our program ($10.5 \le {\rm BHES} \le 16.0$), a total of about 40\%
were shown to possess \metal~$< -1.0$ (85\% of these with \cfe~$> +0.5$, 65\%
with \cfe~$> +1.0$); 21\% with \metal~$< -2.0$ (92\% of these with \cfe~$> +0.5$,
76\% with \cfe~$> +1.0$), and 7\% with \metal~$< -3.0$ (100\% of these with
\cfe~$> +0.5$, 88\% with \cfe~$> +1.0$). The combined sample of this work and
Paper I contains 234 CEMP/metal-poor star candidates with available
medium-resolution spectroscopy, including 108 stars with \metal~$< -1.0$ (74
stars with \cfe~$> +0.5$, 51 with \cfe~$> +1.0$), 57 with \metal~$< -2.0$ (43
stars with \cfe~$> +0.5$, 32 with \cfe~$> +1.0$), and 8 with \metal~$<-3.0$ (8
stars with \cfe~$> +0.5$, 7 with \cfe~$> +1.0$). These numbers are important to
increase the current statistics on \cemp{} stars, since the lack of carbon
abundance data in specific regimes (especially for \metal\ $<-$3.0), limits our
ability to distinguish between statistical and cosmic scatter in the \metal{}
vs. \cfe{} plane. As mentioned in the Introduction, we expect this very
low-metallicity range to be dominated by the CEMP-no class objects, which may
provide important clues to the nucleosynthesis products of the first generations
of stars.

The visual inspection described in Section \ref{visu} also generated a list of
5,288 \cemp{} star candidates to serve as inputs for our continued
medium-resolution spectroscopic follow-up, which aims in particular at
increasing the numbers of known extremely metal-poor stars selected on the basis
of their strong carbon enhancements. The CEMP stars identified in the range
$-2.0 \le$ \metal $\le -1.0$, which we expect to be dominated by the CEMP-s
class, supplement the intermediate-metallicity region, which has been
undersampled by previous follow-up efforts directed primarily at lower
metallicities. 

It is important to recognize that our sample of newly discovered \cemp{} stars is
not suitable for estimation of \cemp{} star fractions in the Galaxy, due to its
selection on carbon in the first place.  There are other, much larger and non
carbon-biased samples, for which this exercise can be carried out.  \citet{carollo2011} 
have done this for the SDSS/SEGUE calibration stars through SDSS DR7.
Their analysis indicates a significant difference in the \cemp{} star fractions
between the inner- and outer-halo components, so a discussion of the {\it global} 
fractions of \cemp{} stars with metallicity no longer appears to be the most
pressing question.  More careful attention will also have to be paid in the
future to the issue of the detectability of the CH G-band feature with
increasing effective temperature, which results in most previous estimates of the
\cemp{} star fractions in reality being lower limits. Studies of the kinematics of
our sample are planned, once the number of stars with available
medium-resolution spectroscopy increases. This information can be used to 
seek assignment of the thick-disk, metal-weak thick-disk, or
inner/outer-halo status for our program stars. Finally, high-resolution
spectroscopic follow-up of our program stars, in particular the VMP and EMP
stars with enhanced carbonicity, will enable their assignment into the
appropriate subclasses of \cemp{} stars.

\acknowledgments V.M.P. acknowledges hospitality
at the Zentrum f\"ur Astronomie der Universit\"at Heidelberg, Landessternwarte,
during which the visual inspection of the candidates took place. V.M.P. and S.R.
acknowledge CNPq, CAPES (PROEX), FAPESP funding (2007/{}04356-3, 2010/{}08996-0)
and JINA. T.C.B. acknowledges partial support for this work from grants AST
07-07776, PHY 02-15783 and PHY 08-22648; Physics Frontier Center/{}Joint
Institute or Nuclear Astrophysics (JINA), awarded by the US National Science
Foundation. N.C. acknowledges support by Sonderforschungsbereich SFB 881
``The Milky Way System'' (subproject A4) of the German Research Foundation (DFG).

These results are based on observations obtained at the
SOAR Telescope (SO2009B-004 / SO2011A-010), 
Gemini North (GN-2011A-Q-88 / GN-2011A-Q-122) 
and Gemini South (GS-2010A-Q-78 / GS-2011A-Q-85 / GS-2011A-Q-86)
Observatories. The Gemini Observatory is operated by the Association 
of Universities for Research in Astronomy, Inc., under a cooperative
agreement with the NSF on behalf the Gemini partnership: the
National Science Foundation (United States), the Science and
Technology Facilities Council (United Kingdom), the National
Research Council (Canada), CONICYT (Chile), the Australian
Research Council (Australia), CNPq (Brazil), and CONICET
(Argentina).

\clearpage
\LongTables

\begin{center}
\begin{deluxetable}{@{}ccccccccc@{}}
\tablecaption{Stellar Data for the Observed Candidates \label{candlist}}
\tablehead{Name & $\alpha$(J2000) & $\delta$(J2000) & BHES & \jk{} & \gpe{}(\AA) & \egp{}(mag) & Telescope & Tag}
\tablewidth{0pt}
\tabletypesize{\small}
\startdata
HE~0002$-$1037 & 00:05:23.0 & $-$10:20:25 & 14.6 & 0.55 & 52.1 & $-$0.41 & SOAR & fhlc\\
HE~0004$-$2546 & 00:06:33.1 & $-$25:29:21 & 11.3 & 0.58 & 18.8 & $-$0.67 & SOAR & mpcc\\
HE~0020$-$2549 & 00:22:39.0 & $-$25:32:58 & 11.6 & 0.66 & 23.9 & $-$0.67 & SOAR & mpcc\\
HE~0046$-$4712 & 00:48:57.0 & $-$46:55:57 & 15.0 & 0.44 & 35.4 & $-$0.47 & SOAR & mpcc\\
HE~0055$-$2507 & 00:57:56.4 & $-$24:51:07 & 11.3 & 0.63 & 22.3 & $-$0.66 & SOAR & mpcc\\
HE~0059$-$6540 & 01:01:18.1 & $-$65:23:56 & 14.6 & 0.56 & 36.8 & $-$0.54 & SOAR & fhlc\\
HE~0113$-$3806 & 01:16:11.8 & $-$37:50:24 & 15.1 & 0.30 & 41.0 & $-$0.54 & SOAR & mpcb\\
HE~0123$+$0023 & 01:26:30.4 & $+$00:39:13 & 15.0 & 0.26 & 29.4 & $-$0.70 & SOAR & unid\\
HE~0134$-$2504 & 01:36:37.3 & $-$24:49:03 & 15.3 & 0.45 & 55.6 & $-$0.30 & SOAR & fhlc\\
HE~0317$-$4705 & 03:18:45.2 & $-$46:54:39 & 13.7 & 0.65 & 39.6 & $-$0.46 & SOAR & fhlc\\
HE~0515$-$3358 & 05:17:08.3 & $-$33:55:03 & 14.6 & 0.62 & 31.6 & $-$0.53 & GEMINI & mpcc\\
HE~0532$-$3819 & 05:33:50.8 & $-$38:17:07 & 15.0 & 0.55 & 38.2 & $-$0.47 & GEMINI & fhlc\\
HE~0546$-$4421 & 05:48:15.6 & $-$44:20:37 & 11.8 & 0.67 & 27.1 & $-$0.60 & SOAR & fhlc\\
HE~0548$-$4508 & 05:50:21.0 & $-$45:07:43 & 12.0 & 0.59 & 35.1 & $-$0.49 & GEMINI & fhlc\\
HE~0854$+$0105 & 08:57:12.6 & $+$00:53:55 & 16.0 & 0.58 & 55.3 & $-$0.28 & GEMINI & fhlc\\
HE~0911$+$0011 & 09:14:16.5 & $+$00:01:08 & 15.4 & 0.36 & 53.1 & $-$0.43 & GEMINI & mpcc\\
HE~0919$-$0049 & 09:22:24.1 & $-$01:02:04 & 16.0 & 0.43 & 48.3 & $-$0.44 & GEMINI & unid\\
HE~0923$-$0016 & 09:26:17.8 & $+$00:29:13 & 15.2 & 0.41 & 38.7 & $-$0.51 & GEMINI & mpcc\\
HE~0927$-$0035 & 09:30:10.8 & $+$00:48:19 & 11.2 & 0.67 & 17.0 & $-$0.68 & SOAR & fhlc\\
HE~0930$-$1047 & 09:33:06.4 & $-$11:00:53 & 15.8 & 0.49 & 46.0 & $-$0.40 & GEMINI & mpcb\\
HE~0932$+$0005 & 09:35:03.5 & $+$00:08:17 & 15.1 & 0.58 & 52.8 & $-$0.37 & SOAR & fhlc\\
HE~0932$-$0838 & 09:34:36.6 & $-$08:52:08 & 15.8 & 0.32 & 47.1 & $-$0.42 & GEMINI & mpcc\\
HE~0942$-$0446 & 09:44:42.2 & $-$05:00:24 & 15.8 & 0.34 & 49.2 & $-$0.41 & GEMINI & mpcc\\
HE~0943$-$0227 & 09:46:14.0 & $-$02:40:57 & 13.6 & 0.37 & 35.1 & $-$0.55 & GEMINI & fhlc\\
HE~0946$-$0737 & 09:48:45.4 & $-$07:51:14 & 15.9 & 0.37 & 45.4 & $-$0.45 & GEMINI & mpcc\\
HE~0954$+$0219 & 09:56:48.9 & $+$02:05:20 & 14.3 & 0.67 & 43.9 & $-$0.46 & SOAR & fhlc\\
HE~0954$-$0744 & 09:56:29.7 & $-$07:59:10 & 15.1 & 0.42 & 31.2 & $-$0.53 & GEMINI & mpcc\\
HE~1006$-$1237 & 10:08:48.2 & $-$12:52:24 & 15.2 & 0.31 & 43.8 & $-$0.58 & GEMINI & unid\\
HE~1007$-$1343 & 10:09:50.2 & $-$13:58:08 & 15.6 & 0.33 & 48.3 & $-$0.47 & GEMINI & unid\\
HE~1013$-$1648 & 10:15:40.6 & $-$17:03:53 & 15.7 & 0.47 & 48.7 & $-$0.52 & GEMINI & unid\\
HE~1016$-$1625 & 10:19:11.6 & $-$16:40:47 & 15.6 & 0.45 & 48.7 & $-$0.48 & GEMINI & mpcb\\
HE~1022$-$0730 & 10:24:39.3 & $-$07:45:59 & 14.9 & 0.37 & 28.9 & $-$0.58 & GEMINI & mpcb\\
HE~1022$-$1621 & 10:24:40.5 & $-$16:36:57 & 15.6 & 0.39 & 46.1 & $-$0.54 & GEMINI & unid\\
HE~1027$-$1217 & 10:29:29.9 & $-$12:32:31 & 15.1 & 0.43 & 32.2 & $-$0.67 & GEMINI & mpcb\\
HE~1029$-$1757 & 10:31:55.3 & $-$18:12:42 & 14.0 & 0.45 & 37.5 & $-$0.46 & GEMINI & fhlc\\
HE~1032$-$1809 & 10:35:18.8 & $-$18:25:01 & 15.5 & 0.67 & 58.0 & $-$0.28 & GEMINI & fhlc\\
HE~1032$-$2042 & 10:35:06.9 & $-$20:57:54 & 15.9 & 0.40 & 46.9 & $-$0.53 & GEMINI & mpcc\\
HE~1034$-$1632 & 10:36:56.4 & $-$16:48:08 & 15.2 & 0.34 & 37.1 & $-$0.68 & GEMINI & mpcc\\
HE~1035$-$1603 & 10:37:38.7 & $-$16:18:44 & 15.6 & 0.59 & 50.5 & $-$0.30 & GEMINI & unid\\
HE~1037$-$0301 & 10:39:40.1 & $-$03:17:08 & 15.8 & 0.48 & 52.5 & $-$0.38 & GEMINI & mpcc\\
HE~1040$-$1957 & 10:42:59.4 & $-$20:12:54 & 15.9 & 0.63 & 48.2 & $-$0.43 & GEMINI & unid\\
HE~1042$-$1107 & 10:44:58.5 & $-$11:23:15 & 16.0 & 0.42 & 45.8 & $-$0.56 & GEMINI & unid\\
HE~1043$-$1516 & 10:45:51.0 & $-$15:32:23 & 15.6 & 0.61 & 53.2 & $-$0.31 & GEMINI & fhlc\\
HE~1045$-$1313 & 10:48:15.3 & $-$13:29:01 & 16.0 & 0.40 & 45.2 & $-$0.36 & GEMINI & unid\\
HE~1046$-$1352 & 10:48:29.9 & $-$14:08:12 & 15.3 & 0.47 & 52.1 & $-$0.41 & GEMINI & mpca\\
HE~1047$-$1140 & 10:49:41.6 & $-$11:56:15 & 15.2 & 0.59 & 45.7 & $-$0.48 & GEMINI & unid\\
HE~1053$-$2017 & 10:56:25.3 & $-$20:33:04 & 15.8 & 0.62 & 55.3 & $-$0.32 & GEMINI & mpcc\\
HE~1054$-$2718 & 10:57:06.7 & $-$27:34:30 & 15.9 & 0.30 & 58.9 & $+$0.08 & GEMINI & mpcc\\
HE~1055$-$2647 & 10:57:29.0 & $-$27:03:50 & 13.8 & 0.69 & 30.4 & $-$0.49 & SOAR & mpcc\\
HE~1106$-$0725 & 11:09:28.6 & $-$07:41:20 & 16.0 & 0.40 & 45.8 & $-$0.54 & GEMINI & mpcb\\
HE~1110$-$2529 & 11:13:04.1 & $-$25:45:59 & 15.6 & 0.68 & 59.8 & $-$0.23 & GEMINI & fhlc\\
HE~1111$-$2817 & 11:14:09.4 & $-$28:33:36 & 16.0 & 0.45 & 45.6 & $-$0.39 & GEMINI & mpcb\\
HE~1111$-$3026 & 11:13:45.0 & $-$30:42:48 & 15.6 & 0.53 & 51.7 & $-$0.32 & GEMINI & fhlc\\
HE~1112$-$1140 & 11:15:14.5 & $-$11:56:50 & 11.4 & 0.65 & 16.2 & $-$0.65 & SOAR & mpcc\\
HE~1124$-$2343 & 11:26:56.2 & $-$23:59:53 & 11.2 & 0.57 & 20.5 & $-$0.68 & SOAR & mpcc\\
HE~1126$-$1229 & 11:28:48.5 & $-$12:46:07 & 10.9 & 0.65 & 15.2 & $-$0.67 & SOAR & fhlc\\
HE~1134$-$1731 & 11:37:11.4 & $-$17:47:45 & 15.4 & 0.54 & 53.6 & $-$0.27 & GEMINI & unid\\
HE~1140$-$2814 & 11:43:03.2 & $-$28:31:02 & 11.8 & 0.61 & 14.5 & $-$0.64 & SOAR & fhlc\\
HE~1141$-$3140 & 11:43:51.7 & $-$31:57:35 & 10.5 & 0.67 & 18.1 & $-$0.65 & SOAR & fhlc\\
HE~1142$-$1058 & 11:44:45.9 & $-$11:14:56 & 14.8 & 0.30 & 45.8 & $-$0.44 & GEMINI & mpcc\\
HE~1144$-$2555 & 11:46:46.7 & $-$26:12:09 & 15.8 & 0.39 & 45.2 & $-$0.51 & GEMINI & mpcc\\
HE~1146$-$1040 & 11:49:24.5 & $-$10:56:41 & 15.0 & 0.50 & 41.2 & $-$0.48 & GEMINI & mpcc\\
HE~1146$-$1128 & 11:48:47.7 & $-$11:44:47 & 14.7 & 0.33 & 32.5 & $-$0.64 & GEMINI & unid\\
HE~1147$-$0415 & 11:50:30.0 & $-$04:32:14 & 15.5 & 0.47 & 43.9 & $-$0.41 & GEMINI & fhlc\\
HE~1150$-$2800 & 11:53:26.2 & $-$28:17:03 & 16.0 & 0.42 & 46.3 & $-$0.38 & GEMINI & fhlc\\
HE~1153$-$2326 & 11:55:58.7 & $-$23:43:03 & 16.0 & 0.34 & 46.9 & $-$0.53 & GEMINI & mpcb\\
HE~1202$-$2732 & 12:05:20.2 & $-$27:48:52 & 15.3 & 0.60 & 54.3 & $-$0.28 & GEMINI & fhlc\\
HE~1216$-$0739 & 12:18:39.6 & $-$07:55:39 & 15.0 & 0.41 & 40.8 & $-$0.49 & GEMINI & unid\\
HE~1233$-$2435 & 12:36:14.9 & $-$24:52:27 & 15.9 & 0.46 & 49.6 & $-$0.56 & GEMINI & unid\\
HE~1254$-$2320 & 12:56:48.4 & $-$23:36:29 & 16.0 & 0.38 & 51.8 & $-$0.45 & GEMINI & unid\\
HE~1304$-$1128 & 13:07:01.8 & $-$11:44:23 & 15.9 & 0.51 & 52.1 & $-$0.36 & GEMINI & mpcc\\
HE~1315$-$2807 & 13:18:34.2 & $-$28:23:02 & 12.5 & 0.65 & 31.4 & $-$0.57 & SOAR & fhlc\\
HE~1328$-$1740 & 13:31:22.8 & $-$17:56:21 & 15.2 & 0.46 & 44.4 & $-$0.34 & GEMINI & mpcc\\
HE~1329$-$2347 & 13:32:03.2 & $-$24:02:57 & 16.0 & 0.60 & 41.3 & $-$0.52 & GEMINI & mpca\\
HE~1336$-$1832 & 13:39:38.0 & $-$18:47:21 & 12.5 & 0.67 & 25.0 & $-$0.56 & SOAR & fhlc\\
HE~1337$-$2608 & 13:39:49.2 & $-$26:24:10 & 15.6 & 0.38 & 37.8 & $-$0.64 & GEMINI & unid\\
HE~1342$-$2731 & 13:44:53.1 & $-$27:46:38 & 11.2 & 0.64 & 17.6 & $-$0.67 & SOAR & mpcc\\
HE~1343$-$0626 & 13:45:53.5 & $-$06:41:28 & 15.5 & 0.42 & 48.3 & $-$0.51 & GEMINI & mpcc\\
HE~1348$-$3057 & 13:51:05.3 & $-$31:12:28 & 14.8 & 0.64 & 60.2 & $-$0.20 & GEMINI & fhlc\\
HE~1350$-$2422 & 13:53:33.9 & $-$24:37:17 & 13.7 & 0.60 & 35.5 & $-$0.45 & SOAR & fhlc\\
HE~1350$-$2734 & 13:53:09.5 & $-$27:49:14 & 14.5 & 0.68 & 36.2 & $-$0.42 & GEMINI & fhlc\\
HE~1401$-$1236 & 14:03:41.9 & $-$12:50:39 & 15.9 & 0.43 & 46.9 & $-$0.44 & GEMINI & mpcc\\
HE~1428$-$1950 & 14:30:59.4 & $-$20:03:42 & 13.1 & 0.62 & 33.8 & $-$0.48 & SOAR & fhlc\\
HE~1444$-$1219 & 14:47:15.6 & $-$12:31:45 & 15.1 & 0.57 & 29.1 & $-$0.64 & SOAR & mpcb\\
HE~1501$-$0858 & 15:03:46.2 & $-$09:10:12 & 12.0 & 0.64 & 20.4 & $-$0.63 & SOAR & mpcc\\
HE~1503$-$0918 & 15:05:50.9 & $-$09:30:24 & 14.8 & 0.35 & 37.1 & $-$0.54 & SOAR & unid\\
HE~1507$-$1122 & 15:10:09.9 & $-$11:33:20 & 10.6 & 0.66 & 17.2 & $-$0.67 & SOAR & fhlc\\
HE~1508$-$0736 & 15:11:15.8 & $-$07:47:30 & 14.9 & 0.28 & 30.2 & $-$0.61 & SOAR & mpcc\\
HE~1509$-$1437 & 15:12:30.3 & $-$14:48:15 & 11.8 & 0.66 & 18.6 & $-$0.61 & SOAR & mpcb\\
HE~1514$-$0943 & 15:17:36.0 & $-$09:53:59 & 15.2 & 0.31 & 35.1 & $-$0.57 & GEMINI & mpcb\\
HE~1516$-$0903 & 15:18:58.4 & $-$09:14:38 & 12.8 & 0.66 & 30.7 & $-$0.55 & SOAR & mpcc\\
HE~1523$-$1155 & 15:26:41.0 & $-$12:05:43 & 14.6 & 0.55 & 52.5 & $-$0.23 & SOAR & fhlc\\
HE~1937$-$6314 & 19:42:00.7 & $-$63:06:57 & 13.6 & 0.52 & 43.0 & $-$0.37 & SOAR & fhlc\\
HE~1939$-$6626 & 19:44:38.8 & $-$66:18:49 & 13.7 & 0.67 & 27.6 & $-$0.46 & SOAR & fhlc\\
HE~2006$-$5334 & 20:10:20.7 & $-$53:25:52 & 15.0 & 0.85 & 40.5 & $-$0.35 & SOAR & unid\\
HE~2030$-$5323 & 20:33:53.3 & $-$53:13:17 & 14.3 & 0.53 & 28.9 & $-$0.60 & SOAR & mpcb\\
HE~2030$-$6056 & 20:34:59.8 & $-$60:45:42 & 12.3 & 0.70 & 24.5 & $-$0.58 & SOAR & mpcc\\
HE~2033$-$6206 & 20:37:44.3 & $-$61:55:43 & 14.7 & 0.32 & 34.9 & $-$0.52 & SOAR & mpcc\\
HE~2043$-$5525 & 20:47:02.1 & $-$55:14:39 & 14.8 & 0.55 & 41.7 & $-$0.47 & SOAR & mpcc\\
HE~2056$-$6128 & 21:00:03.4 & $-$61:17:05 & 15.0 & 0.41 & 34.0 & $-$0.62 & SOAR & mpcc\\
HE~2118$-$5654 & 21:22:19.9 & $-$56:41:12 & 14.9 & 0.47 & 41.0 & $-$0.54 & SOAR & mpcc\\
HE~2121$-$5308 & 21:25:20.6 & $-$52:55:41 & 13.3 & 0.64 & 34.8 & $-$0.49 & SOAR & fhlc\\
HE~2125$-$3447 & 21:28:04.0 & $-$34:33:57 & 13.5 & 0.69 & 27.6 & $-$0.49 & SOAR & fhlc\\
HE~2134$-$0637 & 21:37:01.3 & $-$06:23:39 & 15.0 & 0.35 & 29.8 & $-$0.70 & SOAR & unid\\
HE~2135$-$0759 & 21:38:17.0 & $-$07:46:10 & 11.3 & 0.69 & 13.5 & $-$0.68 & SOAR & fhlc\\
HE~2136$-$5928 & 21:40:08.0 & $-$59:14:31 & 15.2 & 0.46 & 39.1 & $-$0.48 & SOAR & fhlc\\
HE~2138$-$5620 & 21:42:08.5 & $-$56:06:49 & 15.0 & 0.62 & 30.0 & $-$0.63 & SOAR & mpcc\\
HE~2146$-$0247 & 21:49:06.8 & $-$02:33:45 & 15.1 & 0.58 & 30.4 & $-$0.67 & SOAR & mpcc\\
HE~2148$-$1058 & 21:50:44.9 & $-$10:44:49 & 10.7 & 0.65 & 13.9 & $-$0.68 & SOAR & fhlc\\
HE~2151$-$0643 & 21:54:08.6 & $-$06:29:29 & 15.1 & 0.33 & 25.6 & $-$0.70 & SOAR & mpcc\\
HE~2155$-$3750 & 21:58:09.7 & $-$37:36:20 & 14.1 & 0.50 & 40.8 & $-$0.46 & SOAR & fhlc\\
HE~2200$-$1146 & 22:03:05.9 & $-$11:32:07 & 15.1 & 0.35 & 28.7 & $-$0.72 & SOAR & mpcb\\
HE~2211$-$1806 & 22:14:23.9 & $-$17:51:28 & 15.3 & 0.44 & 57.6 & $-$0.29 & SOAR & fhlc\\
HE~2220$-$2250 & 22:22:57.8 & $-$22:35:42 & 11.5 & 0.65 & 14.5 & $-$0.67 & SOAR & mpcc\\
HE~2229$-$1619 & 22:32:35.0 & $-$16:04:15 & 15.1 & 0.52 & 41.7 & $-$0.49 & SOAR & mpcc\\
HE~2324$-$0424 & 23:26:58.9 & $-$04:08:23 & 14.6 & 0.46 & 34.6 & $-$0.64 & SOAR & mpcb\\
HE~2339$-$3236 & 23:42:07.0 & $-$32:19:26 & 14.3 & 0.58 & 32.0 & $-$0.58 & SOAR & fhlc\\
\enddata
\end{deluxetable}
\end{center}

\clearpage

\begin{center}
\begin{deluxetable}{crcccccc}
\tablecaption{Atmospheric Parameters and Carbonicity Estimates
for the Observed Candidates. \label{atmpar}}
\tablehead{Name & V (km/{}s) & \teff{} (K) & \logg{} (cgs) & \metal{} & \cfe{}}
\tablewidth{0pt}
\tabletypesize{\small}
\startdata
HE~0002$-$1037 & $-$73.0 & 4895 & 3.09 & $-$3.07 & $+$1.12\\
HE~0004$-$2546 & 31.2 & 4916 & 4.01 & $-$0.59 & $+$0.18\\
HE~0008$+$0049 & $-$12.4 & 4862 & 3.56 & $-$1.27 & $-$0.32\\
HE~0020$-$2549 & 25.6 & 5078 & 2.74 & $-$0.91 & $-$0.03\\
HE~0024$-$0550 & 17.8 & 5538 & 2.98 & $-$1.78 & $+$0.31\\
HE~0034$-$0011 & $-$218.8 & 5934 & 2.52 & $-$1.74 & $+$1.71\\
HE~0035$-$5803 & 30.6 & 5807 & 3.75 & $-$0.70 & $+$0.47\\
HE~0046$-$4712 & $-$119.0 & 5340 & 3.62 & $-$1.04 & $+$0.63\\
HE~0053$-$0356 & $-$86.2 & 5752 & 2.21 & $-$1.84 & $+$1.85\\
HE~0055$-$2507 & 35.9 & 4895 & 3.29 & $-$0.34 & $+$0.09\\
HE~0058$+$0141 & $-$48.4 & 6271 & 4.08 & $-$0.68 & $+$0.48\\
HE~0059$-$6540 & $-$28.4 & 4727 & 1.00 & $-$3.17 & $+$1.20\\
HE~0100$-$4957 & 126.9 & 5970 & 3.54 & $-$1.03 & $+$0.28\\
HE~0102$-$0004 & $-$191.6 & 5814 & 3.45 & $-$2.45 & $+$0.64\\
HE~0113$-$3806 & $-$131.2 & 5886 & 3.14 & $-$2.08 & $+$1.90\\
HE~0118$-$4834 & $-$144.2 & 5609 & 2.27 & $-$2.42 & $+$2.14\\
HE~0123$+$0023 & $-$304.0 & 6322 & 2.97 & $-$1.81 & $+$2.05\\
HE~0134$-$2504 & $-$41.5 & 5150 & 2.42 & $-$3.10 & $+$1.85\\
HE~0156$-$5608 & 213.7 & 5137 & 4.27 & $-$2.00 & $+$0.47\\
HE~0159$-$5216 & $-$18.1 & 5117 & 1.85 & $-$2.03 & $+$0.78\\
HE~0214$-$0818 & $-$48.6 & 6177 & 2.85 & $-$1.02 & $+$1.32\\
HE~0307$-$5339 & 163.0 & 5465 & 2.45 & $-$2.14 & $+$1.34\\
HE~0316$-$2903 & 198.2 & 5795 & 3.15 & $-$1.56 & $+$1.35\\
HE~0317$-$4705 & 173.5 & 4231 & 3.06 & $-$3.26 & $+$0.95\\
HE~0320$-$1242 & 44.6 & 5749 & 4.17 & $-$0.43 & $+$0.18\\
HE~0322$-$3720 & $-$38.5 & 4660 & 4.55 & ~~0.00 & $+$0.02\\
HE~0336$-$3948 & 101.4 & 5952 & 3.86 & $-$0.53 & $+$0.28\\
HE~0340$-$3933 & $-$67.1 & 5991 & 4.03 & $-$0.04 & $+$0.01\\
HE~0345$+$0006 & $-$51.2 & 5199 & 2.61 & $-$2.74 & $+$0.49\\
HE~0405$-$4411 & 45.0 & 6180 & 3.55 & $-$1.23 & $+$0.85\\
HE~0414$-$4645 & 38.5 & 5820 & 3.78 & $-$1.08 & $+$0.33\\
HE~0440$-$5525 & 26.2 & 6361 & 3.65 & $-$0.52 & $+$0.47\\
HE~0444$-$3536 & 117.0 & 5197 & 2.42 & $-$1.66 & $+$0.96\\
HE~0449$-$1617 & 70.6 & 5766 & 4.42 & $-$0.30 & $-$0.14\\
HE~0451$-$3127 & 305.2 & 5505 & 1.85 & $-$2.43 & $+$1.49\\
HE~0500$-$5603 & 99.6 & 5808 & 3.67 & $-$0.56 & $+$0.19\\
HE~0509$-$1611 & 92.9 & 5108 & 3.05 & $-$0.49 & $-$0.01\\
HE~0511$-$3411 & 50.6 & 5873 & 4.04 & $-$0.38 & $+$0.16\\
HE~0514$-$5449 & 119.2 & 6505 & 4.19 & $-$0.55 & $+$0.28\\
HE~0515$-$3358 & 16.6 & 5057 & 1.99 & $-$0.60 & $+$0.26\\
HE~0518$-$3941 & 18.4 & 6562 & 3.36 & $-$0.56 & $+$0.58\\
HE~0532$-$3819 & $-$40.6 & 5031 & 1.69 & $-$1.48 & $+$0.94\\
HE~0535$-$4842 & 32.1 & 5935 & 4.30 & $-$0.58 & $+$0.25\\
HE~0536$-$5647 & 89.5 & 5639 & 4.08 & $-$0.46 & $-$0.16\\
HE~0537$-$4849 & 30.9 & 5082 & 4.73 & $-$0.38 & $+$0.13\\
HE~0546$-$4421 & 11.5 & 4966 & 4.87 & ~~0.00 & $-$0.17\\
HE~0548$-$4508 & 51.1 & 5163 & 1.05 & $-$2.42 & $+$2.67\\
HE~0854$+$0105 & $-$13.7 & 4951 & 2.10 & $-$0.72 & $-$0.03\\
HE~0901$-$0003 & $-$4.0 & 5077 & 4.14 & $-$0.26 & $+$0.01\\
HE~0910$-$0126 & 112.3 & 6409 & 3.24 & $-$2.00 & $+$0.92\\
HE~0911$+$0011 & $-$64.1 & 5554 & 4.04 & $-$0.73 & $+$0.05\\
HE~0912$+$0200 & 78.4 & 5003 & 4.12 & $-$0.06 & $+$0.07\\
HE~0918$-$0156 & 65.9 & 4114 & 2.34 & $-$0.99 & $-$0.01\\
HE~0919$-$0049 & $-$11.9 & 5609 & 3.88 & $-$0.35 & $+$0.10\\
HE~0923$-$0016 & $-$10.9 & 5929 & 3.49 & $-$1.27 & $+$0.98\\
HE~0923$-$0323 & 57.7 & 5576 & 3.67 & $-$0.74 & $+$0.28\\
HE~0927$-$0035 & 86.4 & 4737 & 3.09 & $-$0.43 & $-$0.03\\
HE~0928$+$0059 & $-$61.3 & 6482 & 3.67 & $-$0.60 & $+$0.54\\
HE~0930$-$1047 & $-$17.7 & 5387 & 4.14 & $-$1.06 & $+$0.15\\
HE~0932$+$0005 & $-$4.1 & 4816 & 4.81 & $-$0.86 & $+$0.11\\
HE~0932$-$0838 & $-$44.9 & 5648 & 3.10 & $-$0.56 & $+$0.35\\
HE~0933$-$0733 & 20.2 & 5444 & 4.10 & $-$1.04 & $+$0.21\\
HE~0942$-$0446 & $-$70.8 & 5919 & 3.91 & $-$0.07 & $+$0.03\\
HE~0943$-$0227 & 11.5 & 5833 & 3.32 & $-$0.37 & $+$0.51\\
HE~0946$-$0737 & $-$64.5 & 5433 & 4.41 & $-$0.67 & $+$0.12\\
HE~0948$+$0107 & 442.5 & 5081 & 3.88 & $-$2.27 & $-$0.23\\
HE~0948$-$0234 & 114.5 & 5807 & 4.07 & $-$1.02 & $+$0.41\\
HE~0950$-$0401 & 90.4 & 5849 & 3.62 & $-$1.57 & $+$1.32\\
HE~0950$-$1248 & 38.3 & 5554 & 4.01 & $-$0.03 & $-$0.17\\
HE~0954$+$0219 & 237.3 & 5210 & 1.70 & $-$1.87 & $+$1.20\\
HE~0954$-$0744 & $-$55.5 & 5842 & 3.77 & $-$0.24 & $-$0.02\\
HE~1001$-$1621 & $-$36.9 & 6112 & 4.19 & $-$0.46 & $+$0.21\\
HE~1002$-$1405 & 40.2 & 5653 & 3.77 & $-$0.40 & $+$0.28\\
HE~1006$-$1237 & $-$44.3 & 6315 & 3.58 & $-$0.44 & $+$0.49\\
HE~1007$-$1343 & $-$129.2 & 5840 & 4.12 & $-$0.24 & $+$0.05\\
HE~1007$-$1524 & 42.1 & 5782 & 3.36 & $-$0.52 & $+$0.27\\
HE~1009$-$1613 & 28.0 & 5765 & 3.90 & ~~0.00 & $-$0.01\\
HE~1009$-$1646 & $-$34.4 & 5818 & 4.08 & $-$0.52 & $+$0.15\\
HE~1010$-$1445 & 172.2 & 4923 & 3.31 & $-$0.84 & ~~0.00\\
HE~1013$-$1648 & $-$68.8 & 5318 & 3.08 & $-$0.52 & $+$0.27\\
HE~1016$-$1625 & $-$93.9 & 5188 & 3.31 & $-$0.64 & $-$0.04\\
HE~1022$-$0730 & $-$26.0 & 5809 & 3.73 & $-$1.34 & $+$0.59\\
HE~1022$-$1621 & $-$42.1 & 5328 & 4.09 & $-$0.60 & $+$0.02\\
HE~1027$-$1217 & 5.0 & 7505 & 3.75 & $-$0.61 & $+$2.10\\
HE~1029$-$1757 & $-$76.2 & 5696 & 1.73 & $-$3.10 & $+$2.90\\
HE~1032$-$1809 & $-$75.5 & 4505 & 4.76 & $-$0.80 & $+$0.39\\
HE~1032$-$2042 & 45.6 & 5665 & 3.73 & $-$0.35 & $+$0.10\\
HE~1034$-$1632 & 81.1 & 6511 & 3.26 & $-$1.33 & $+$1.26\\
HE~1035$-$1603 & $-$75.8 & 4889 & 4.88 & $-$0.42 & $+$0.10\\
HE~1037$-$0301 & $-$85.9 & 5616 & 3.98 & $-$0.39 & $+$0.05\\
HE~1039$-$1019 & 37.1 & 5519 & 4.14 & $-$0.52 & $+$0.11\\
HE~1040$-$1957 & $-$28.1 & 4513 & 4.43 & $-$0.59 & $+$0.91\\
HE~1042$-$1107 & 12.5 & 5705 & 4.45 & $-$0.63 & $+$0.01\\
HE~1043$-$1516 & $-$93.9 & 4512 & 4.72 & $-$0.81 & $+$0.33\\
HE~1045$+$0226 & 183.4 & 4938 & 1.29 & $-$2.75 & $+$1.95\\
HE~1045$-$1313 & $-$31.3 & 5248 & 3.28 & $-$0.38 & $+$0.10\\
HE~1046$-$1352 & 10.7 & 5233 & 1.81 & $-$3.57 & $+$3.65\\
HE~1046$-$1644 & $-$144.0 & 4915 & 3.64 & $-$0.15 & $-$0.07\\
HE~1047$-$1140 & $-$58.9 & 4954 & 4.44 & $-$0.03 & $-$0.08\\
HE~1049$-$0922 & 1.5 & 4410 & 4.61 & $-$0.26 & $+$0.53\\
HE~1053$-$2017 & $-$72.6 & 5065 & 4.06 & $-$0.43 & $-$0.02\\
HE~1054$-$2718 & $-$80.8 & 5515 & 4.21 & $-$0.12 & $-$0.17\\
HE~1055$-$2647 & 57.8 & 4627 & 2.87 & $-$0.68 & $+$0.03\\
HE~1104$-$0238 & 116.0 & 4075 & 2.08 & $-$1.28 & $-$0.13\\
HE~1106$-$0725 & 133.0 & 5657 & 3.75 & $-$1.32 & $+$0.15\\
HE~1110$-$1625 & 49.0 & 5604 & 4.50 & $-$0.10 & $-$0.11\\
HE~1110$-$2529 & $-$82.2 & 4426 & 4.77 & $-$1.10 & $+$0.25\\
HE~1111$-$2817 & $-$37.8 & 5375 & 4.05 & $-$0.26 & $+$0.01\\
HE~1111$-$3026 & 157.9 & 5035 & 1.23 & $-$2.02 & $+$1.04\\
HE~1112$-$1140 & 23.9 & 4778 & 3.64 & $-$0.48 & $+$0.08\\
HE~1124$-$2343 & 71.2 & 5025 & 4.95 & $-$0.68 & $+$0.13\\
HE~1126$-$1229 & 87.8 & 4693 & 3.13 & $-$0.80 & $+$0.05\\
HE~1129$-$1405 & 170.7 & 5129 & 1.69 & $-$2.29 & $+$0.65\\
HE~1132$-$0915 & 28.3 & 7194 & 4.13 & $-$0.92 & $+$0.81\\
HE~1133$-$0802 & $-$62.8 & 5310 & 3.41 & $-$0.53 & $+$0.22\\
HE~1134$-$1731 & 74.6 & 4831 & 1.74 & $-$0.88 & $+$0.08\\
HE~1135$-$0800 & 112.4 & 5175 & 2.43 & $-$2.17 & $+$0.17\\
HE~1137$-$1259 & 73.3 & 4689 & 4.49 & $-$0.60 & $+$0.23\\
HE~1140$-$2814 & 76.7 & 4650 & 2.56 & $-$0.68 & $-$0.08\\
HE~1141$-$3140 & $-$139.8 & 4688 & 3.78 & $-$0.44 & $+$0.19\\
HE~1142$-$0637 & 97.8 & 5044 & 3.55 & $-$0.94 & $-$0.29\\
HE~1142$-$1058 & 16.8 & 5541 & 4.08 & $-$0.59 & $+$0.30\\
HE~1144$-$2555 & $-$68.1 & 5522 & 3.89 & $-$0.21 & $-$0.08\\
HE~1146$-$1040 & $-$55.5 & 5449 & 4.49 & $-$1.48 & $+$0.35\\
HE~1146$-$1126 & 247.4 & 5041 & 2.07 & $-$2.27 & $+$0.38\\
HE~1146$-$1128 & $-$62.0 & 5915 & 3.42 & $-$0.66 & $+$0.38\\
HE~1147$-$0415 & $-$50.3 & 5042 & 1.05 & $-$2.71 & $+$2.58\\
HE~1147$-$1057 & 63.0 & 5792 & 4.23 & $-$0.43 & $+$0.18\\
HE~1148$-$1020 & 189.7 & 5994 & 3.89 & $-$0.80 & $+$0.34\\
HE~1148$-$1025 & 101.9 & 5718 & 4.22 & $-$0.60 & $+$0.14\\
HE~1150$-$2800 & 0.1 & 5431 & 1.70 & $-$1.78 & $+$1.53\\
HE~1153$-$2326 & 22.6 & 6133 & 2.79 & $-$1.80 & $+$1.75\\
HE~1202$-$2732 & $-$77.7 & 4700 & 4.81 & $-$0.73 & $+$1.01\\
HE~1212$-$1123 & 63.8 & 6167 & 3.67 & $-$1.30 & $+$0.55\\
HE~1216$-$0739 & $-$53.9 & 5335 & 3.77 & $-$0.26 & $+$0.01\\
HE~1217$-$1054 & 26.0 & 4714 & 4.44 & $-$0.34 & $+$0.09\\
HE~1217$-$1633 & 109.1 & 4546 & 1.32 & $-$2.36 & $+$0.61\\
HE~1222$-$1631 & 20.5 & 5062 & 2.10 & $-$2.09 & $+$0.31\\
HE~1223$-$0930 & 127.5 & 4940 & 1.11 & $-$2.52 & $+$2.48\\
HE~1224$-$0723 & 0.2 & 5355 & 4.34 & $-$0.24 & $-$0.02\\
HE~1224$-$1043 & 232.3 & 6227 & 3.44 & $-$1.53 & $+$0.56\\
HE~1228$-$0750 & 273.9 & 5970 & 2.55 & $-$1.52 & $-$0.15\\
HE~1228$-$1438 & 128.6 & 4070 & 2.02 & $-$1.28 & $-$0.12\\
HE~1231$-$3136 & 55.7 & 6410 & 3.38 & $-$1.15 & $+$0.93\\
HE~1233$-$2435 & 15.6 & 5612 & 2.37 & $-$1.65 & $+$1.15\\
HE~1254$-$2320 & $-$73.4 & 5646 & 3.85 & $-$0.19 & $-$0.04\\
HE~1255$-$2734 & $-$64.7 & 5446 & 2.47 & $-$2.32 & $+$1.38\\
HE~1301$+$0014 & $-$8.7 & 5259 & 2.85 & $-$2.55 & $+$0.23\\
HE~1301$-$1405 & $-$9.4 & 6059 & 3.91 & $-$0.52 & $+$0.27\\
HE~1302$-$0954 & 73.5 & 5193 & 2.13 & $-$2.42 & $+$1.05\\
HE~1304$-$1128 & $-$103.7 & 5269 & 4.24 & $-$0.45 & $+$0.04\\
HE~1311$-$3002 & 141.2 & 4783 & 1.06 & $-$2.60 & $+$0.94\\
HE~1315$-$2807 & 118.8 & 5530 & 2.46 & ~~0.00 & $-$0.07\\
HE~1320$-$1130 & 178.7 & 6148 & 3.21 & $-$1.85 & $+$1.78\\
HE~1320$-$1641 & 43.0 & 4051 & 2.25 & $-$0.72 & $+$0.97\\
HE~1321$-$1652 & 59.6 & 5684 & 2.51 & $-$2.41 & $+$2.16\\
HE~1328$-$1740 & $-$77.1 & 5267 & 4.35 & ~~0.00 & $-$0.21\\
HE~1329$-$2347 & $-$104.0 & 4853 & 3.27 & $-$0.38 & $+$0.06\\
HE~1336$-$1832 & $-$28.1 & 6270 & 3.38 & $-$0.08 & $+$0.63\\
HE~1337$-$2608 & $-$62.8 & 5850 & 3.09 & $-$2.62 & $+$2.25\\
HE~1342$-$2731 & $-$45.0 & 4824 & 3.21 & $-$0.39 & $+$0.09\\
HE~1343$+$0137 & 31.9 & 6273 & 3.44 & $-$1.49 & $+$0.58\\
HE~1343$-$0626 & $-$152.8 & 5532 & 1.51 & $-$2.09 & $+$1.65\\
HE~1348$-$3057 & 77.7 & 4964 & 1.08 & $-$1.91 & $+$1.16\\
HE~1350$-$2422 & 224.7 & 5412 & 1.57 & $-$1.56 & $+$1.81\\
HE~1350$-$2734 & $-$59.4 & 4373 & 2.45 & $-$0.89 & $+$0.03\\
HE~1401$-$1236 & 1.0 & 5565 & 1.71 & $-$2.12 & $+$1.87\\
HE~1408$-$0444 & 71.3 & 6622 & 3.66 & $-$2.72 & $+$0.97\\
HE~1409$-$1134 & 14.5 & 5867 & 4.38 & $-$0.02 & $-$0.02\\
HE~1410$-$0549 & 14.6 & 7604 & 4.17 & $-$0.31 & $+$1.04\\
HE~1414$-$1644 & 12.2 & 5176 & 2.62 & $-$2.43 & $+$0.49\\
HE~1418$-$1634 & 61.3 & 5237 & 2.18 & $-$2.20 & $+$0.45\\
HE~1428$-$0851 & $-$12.0 & 5945 & 2.31 & $-$2.04 & $+$0.03\\
HE~1428$-$1950 & 6.0 & 4562 & 3.75 & $-$2.07 & $+$0.49\\
HE~1430$-$1518 & 332.1 & 4399 & 1.52 & $-$1.64 & $+$0.29\\
HE~1444$-$1219 & $-$65.2 & 5076 & 2.49 & $-$2.49 & $+$0.24\\
HE~1447$-$1533 & 12.6 & 4181 & 2.44 & $-$0.59 & $+$0.01\\
HE~1448$-$1406 & $-$253.5 & 6467 & 3.09 & $-$1.37 & $+$0.15\\
HE~1458$-$0923 & $-$401.7 & 5473 & 1.76 & $-$2.32 & $+$2.51\\
HE~1458$-$1022 & $-$146.3 & 5025 & 2.07 & $-$2.26 & $+$0.42\\
HE~1458$-$1226 & $-$59.9 & 5122 & 4.55 & $-$0.27 & $+$0.22\\
HE~1501$-$0858 & $-$155.1 & 4900 & 4.43 & $-$0.17 & $-$0.08\\
HE~1503$-$0918 & $-$56.4 & 5856 & 3.48 & $-$1.03 & $+$0.78\\
HE~1504$-$1534 & $-$12.3 & 4028 & 2.20 & $-$0.57 & $+$0.04\\
HE~1505$-$0826 & 12.6 & 6513 & 3.94 & $-$0.53 & $+$0.49\\
HE~1507$-$1055 & 118.4 & 4178 & 2.18 & $-$1.66 & $-$0.03\\
HE~1507$-$1104 & 53.0 & 4056 & 2.26 & $-$0.65 & $+$0.15\\
HE~1507$-$1122 & $-$107.8 & 4404 & 2.53 & $-$0.92 & $-$0.21\\
HE~1508$-$0736 & 268.4 & 5897 & 4.19 & $-$0.62 & $-$0.26\\
HE~1509$-$1437 & $-$130.7 & 4423 & 3.06 & $-$0.74 & $+$1.00\\
HE~1514$-$0943 & $-$127.4 & 6845 & 3.49 & $-$1.46 & $+$2.31\\
HE~1516$-$0107 & $-$55.9 & 5275 & 2.67 & $-$2.11 & $+$0.19\\
HE~1516$-$0903 & $-$146.8 & 4898 & 4.79 & $-$0.19 & $-$0.06\\
HE~1518$-$0541 & $-$20.3 & 4808 & 4.98 & $-$0.63 & $+$0.31\\
HE~1523$-$1155 & $-$132.7 & 4741 & 1.36 & $-$2.83 & $+$0.94\\
HE~1527$-$0740 & $-$39.8 & 6264 & 2.97 & $-$2.08 & $+$1.13\\
HE~1529$-$0838 & $-$50.4 & 5744 & 4.32 & $-$0.38 & $+$0.12\\
HE~1937$-$6314 & 148.6 & 4516 & 2.26 & $-$3.80 & $+$2.90\\
HE~1939$-$6626 & $-$153.1 & 4760 & 3.72 & ~~0.00 & $-$0.16\\
HE~2006$-$5334 & $-$172.3 & 4105 & 2.52 & $-$0.89 & $-$0.02\\
HE~2025$-$5221 & 148.7 & 5619 & 2.27 & $-$2.32 & $+$2.55\\
HE~2030$-$5323 & 153.1 & 5090 & 1.79 & $-$2.12 & $+$0.85\\
HE~2030$-$6056 & $-$43.5 & 4718 & 3.43 & $-$0.28 & $+$0.03\\
HE~2033$-$6206 & $-$85.6 & 5956 & 3.35 & $-$1.31 & $+$1.09\\
HE~2043$-$5525 & 61.6 & 4870 & 1.34 & $-$2.65 & $+$1.13\\
HE~2052$-$5610 & 207.4 & 5840 & 2.32 & $-$2.15 & $+$2.46\\
HE~2056$-$6128 & $-$224.4 & 5204 & 3.99 & $-$1.32 & $+$0.07\\
HE~2112$-$5236 & 221.6 & 4996 & 1.33 & $-$1.96 & $+$1.21\\
HE~2118$-$5654 & $-$14.9 & 5261 & 1.96 & $-$2.71 & $+$1.79\\
HE~2121$-$5308 & $-$72.1 & 4464 & 1.05 & $-$2.75 & $+$0.50\\
HE~2125$-$3447 & $-$133.7 & 4468 & 2.97 & $-$0.40 & $+$0.15\\
HE~2134$-$0637 & $-$161.2 & 6220 & 2.95 & $-$2.32 & $+$2.63\\
HE~2135$-$0759 & 31.6 & 4742 & 3.57 & $-$0.27 & $+$0.02\\
HE~2136$-$5928 & $-$90.5 & 5163 & 1.15 & $-$1.85 & $+$1.13\\
HE~2138$-$5620 & $-$129.2 & 5435 & 4.07 & $-$0.64 & $-$0.21\\
HE~2140$-$4746 & 14.7 & 5958 & 3.66 & $-$1.16 & $+$0.28\\
HE~2146$-$0247 & $-$240.7 & 5516 & 3.83 & $-$0.31 & $-$0.19\\
HE~2148$-$1058 & 39.6 & 4966 & 2.93 & $-$0.19 & $-$0.23\\
HE~2151$-$0332 & $-$143.8 & 5259 & 1.62 & $-$2.51 & $+$1.53\\
HE~2151$-$0643 & $-$68.6 & 5855 & 4.27 & $-$0.69 & $+$0.17\\
HE~2155$-$3750 & 27.7 & 4858 & 1.13 & $-$2.85 & $+$1.60\\
HE~2200$-$1146 & $-$6.0 & 6032 & 4.09 & $-$0.37 & $+$0.03\\
HE~2201$-$1108 & $-$154.6 & 6111 & 3.87 & $-$1.39 & $+$0.48\\
HE~2207$-$0912 & $-$115.5 & 5347 & 2.87 & $-$2.48 & $+$0.43\\
HE~2209$-$1212 & 72.5 & 6193 & 3.71 & $-$0.18 & $+$0.20\\
HE~2211$-$1806 & 46.3 & 4782 & 1.24 & $-$3.38 & $+$1.32\\
HE~2219$-$1357 & 80.9 & 7091 & 3.80 & $-$0.49 & $+$0.74\\
HE~2220$-$2250 & 0.2 & 4798 & 2.86 & $-$0.07 & $-$0.36\\
HE~2229$-$1619 & $-$253.4 & 4956 & 1.58 & $-$2.51 & $+$1.38\\
HE~2231$-$0710 & 55.4 & 5200 & 2.48 & $-$0.71 & $-$0.04\\
HE~2257$-$5710 & $-$3.4 & 5107 & 2.02 & $-$2.71 & $+$0.96\\
HE~2324$-$0424 & $-$125.9 & 5261 & 3.96 & $-$0.37 & $-$0.24\\
HE~2339$-$3236 & $-$146.2 & 4643 & 4.80 & $-$0.61 & $+$0.91\\
HE~2353$-$5329 & 48.5 & 6013 & 2.81 & $-$1.85 & $+$2.10\\
\enddata

\end{deluxetable}
\end{center}

\appendix

\section{The list of Candidate Metal-poor Stars}
\label{aplist}

Table A.1 lists all 5,288 candidate metal-poor stars selected by means of the
visual inspection on the \hes{} plates, excluding the classes {\emph{art}},
{\emph{nois}} and {\emph{ovl}}. The table is made available electronically only.
The columns contain the following information:

\begin{tabular}{ll}

hename    & HE designation \\
HESid     & Unique HES identifier \\
ra2000    & Right ascension at equinox 2000.0 \\
dec2000   & Declination at equinox 2000.0 \\
objtype   & Object type ({\emph{stars / bright}}) \\
BHES      & Photographic $B$ magnitude \\
BVmphs    & \bvz \\
JminK0    & \jk \\
KPHES     & KP index, measured in \hes{} spectrum \\
GPE       & GPE index, measured in \hes{} spectrum \\
EGP       & EGP index, measured in \hes{} spectrum \\
canclass  & Candidate class (see Table \ref{canclass}) \\

\end{tabular}

\end{document}